\documentclass[11pt,a4paper]{scrartcl}

\usepackage{ILD}

\usepackage[symbol]{footmisc}
\usepackage{feynmf}
\usepackage{textpos}
\usepackage{lineno}
\usepackage[]{units}
\usepackage{bm} 
\usepackage{float} 
\usepackage{makecell} 
\usepackage{tabularx}
\usepackage{multicol}
\usepackage{multirow}

\usepackage{subcaption}

\title{Kinematic Edge Detection Using Finite Impulse Response Filters}

\date{\today}

\addauthor{M. Berggren, S. Caiazza, M. Chera, J. List}{\institute{1}}

\addinstitute{1}{DESY}

\abstract{Various physics observables can be determined from the localisation of distinct edge-like features in distributions of measurement values. In this paper, we address the observation that neither differentiating nor fitting the measured distributions is robust against significant fluctuations in the experimental data. We propose the application of Finite Impulse Response (FIR) filters instead. To demonstrate the method, we consider the typical case in particle physics in which the precise localisation of kinematic edges, often blurred by e.g.\ background contributions and detector effects, is crucial for determining particle masses. We show that even for binned data, typical for high energy physics, the optimal FIR filter kernel can be approximated by the {\em first derivative of a Gaussian} (FDOG). We study two highly complementary supersymmetric scenarios that, if  realised in nature, could be observed at a future high-energy $e^+e^-$ collider such as the International Linear Collider (ILC) or the Compact Linear Collider (CLIC). The first scenario considers the production of $\tilde{e}^{\pm}-$pairs while the second focuses on the $\tilde{\chi}^{\pm}_{1}$ and $\tilde{\chi}^{0}_{2}$-pair production. We demonstrate that the FIR filter method for edge extraction is superior to previously employed methods in terms of robustness and precision.
	
}

\titlecomment{This work was carried out in the framework of the ILD  concept group}

\graphicspath{ {.}}

\begin{document}

\titlepage

  \section{Introduction}

Physics observables are commonly extracted by detecting the position of differential features in distributions of experimental data. An important example in particle physics is the measurement of edge positions in kinematic distributions, which can reveal the kinematic end-points of a process. In an actual experiment, such an edge position will be subject to noise from statistical fluctuations of the signal process as well as from remaining background contributions, and will be blurred e.g.\ by the finite resolution of the detector. 

Various techniques can be used to extract the sought after kinematic feature from such distributions. The most commonly used technique is to fit the distribution with a differentiable function that can model the feature to be measured such that the relevant parameter value(s) can be extracted from the fitted function. This technique implies a model choice, i.e.\ an assumption regarding the analytical form of the feature to detect. In many cases, especially when there are detector effects and background processes affecting the distribution, there is no unique, model-independent choice for an analytic description. It might be also unfeasible to find any viable analytical function that can model the distribution such that it stays robust against small changes of its shape.

In this article we describe a computational method based on finite impulse response filters. Related techniques are not uncommon in digital image processing, and are also used elsewhere, e.g.\ in the automatic calibration system of the LIGO experiment~\cite{Viets:2017yvy}. Here, we adjust the method to the application of typical particle physics distributions. 
As an example, we focus on the measurement of kinematic end-points of physical processes which are particularly relevant for the physics program of future electron-positron colliders. Two complementary study cases have been chosen. The first one exhibits a large signal with a sharp edge over a small background. Here, we will especially study the systematic uncertainties of the filter method. The second example has a much smaller signal-to-background ratio and a shallow edge, making a robust determination of its position challenging.

This paper is structured as follows: in Sec.~\ref{sec:edgedetect} we describe and characterize our edge detection method. The two study cases are introduced  in Sec.~\ref{sec:studycases}, while the application of the algorithm to the two benchmarks is discussed and compared to alternative approaches in Sec.~\ref{sec:selectrons} and~\ref{sec:gauginos}, respectively.
We conclude by summarizing our results in Sec.~\ref{sec:conclusions}.

  \section{A computational approach to edge detection}
\label{sec:edgedetect}

In the general case of a continuous function $\bm{g(x)}$, an edge-like feature is characterised by a local maximum or minimum in the first derivative $\bm{g'(x)}$ and a zero-crossing in the second derivative $\bm{g''(x)}$. Unfortunately, in the case of experimental data, the physical signal is disturbed by noise and small variations in the noise amplitude can have large impact on the results of the differential operator~\cite{Torre1986}. One possibility to regularise this problem is to convolute the target function $\bm{g(x)}$ with an appropriate function $\bm{f(x)}$ \emph{before} the differentiation stage. 

Assuming $\bm{f}$ is differentiable and absolutely integrable, and $\bm{g}$ is absolutely integrable, the derivative of the convolution (*) of these two functions, denoted here as the response $\bm{R}$, can be written as

\begin{align}
R(x) = \frac{d}{dx}(f(x) * g(x)) = \frac{df(x)}{dx} * g(x) = h(x) * g(x) .
\end{align}

The edge-like features in $\bm{g(x)}$ can be identified from the maxima of $\bm{h(x) * g(x)}$ or the zero-crossings of $\bm{h'(x) * g(x)}$. In this context, the function $\bm{df(x)/dx =  h(x)}$ is called \emph{a matched filter} and can be characterised by the result of its convolution with a delta function. The convolution result is called the algorithm's \emph{Impulse Response} (IR). Since the filter unambiguously defines its impulse response we will use the two terms as synonyms. When the filter values drop to 0 everywhere outside a finite range in its domain, that function is known as a \emph{Finite Impulse Response} (FIR) filter, otherwise it is called an \emph{Infinite Impulse Response} (IIR) filter.

In the analysis of physics data though, the typical observable distributions are discrete, therefore the filters and the convolution algorithm should be defined in the same domain. For this discussion, we assume that our input distributions are defined on a regular grid, i.e. that the distance between the samples in the distribution domain is constant, and that the error associated to each sample is independent from all the others. A histogram with constant binning is a special case of such a distribution. Consequently, the filter can also be represented by a discrete function on the same regular grid. Considering the input function $\bm{g_d(i)}$ and a filter represented by a discrete function $\bm{h_d(k)}$, the \emph{Discrete Response Function} $\bm{R_d(i)}$, that is their discrete convolution, can be written as:

\begin{align}
R_d(i) = \sum_{k = -N}^{N} h_d(k)\cdot g_d(i-k)
\label{discr_response}
\end{align}

where N is the value after which the filter amplitude drops to 0. 

In the case of a continuous distribution, the optimal characteristics of a differential filter were described in Ref.~\cite{Canny:4767851}, which showed that the optimal differential filter for a continuous distribution disturbed by random noise, called the \emph{Canny filter}, is very similar to the \emph{First Derivative of a Gaussian} (FDOG), which will be defined in Sec.~\ref{sec:FilterDefinitions}. In the case of a discrete distribution the results of the optimization can differ substantially, as described extensively in \cite{Demigny:1997:DEC:271437.271439}, and different types of filters may be used to obtain the optimal results. Therefore we will redefine those optimization criteria for the specific case of the detection of an edge-like feature in a one-dimensional distribution and develop an operational procedure to evaluate the algorithm performance to optimize the parameters of its operation in the following subsections.

\begin{figure}
	\centering
	\includegraphics[width=\textwidth]{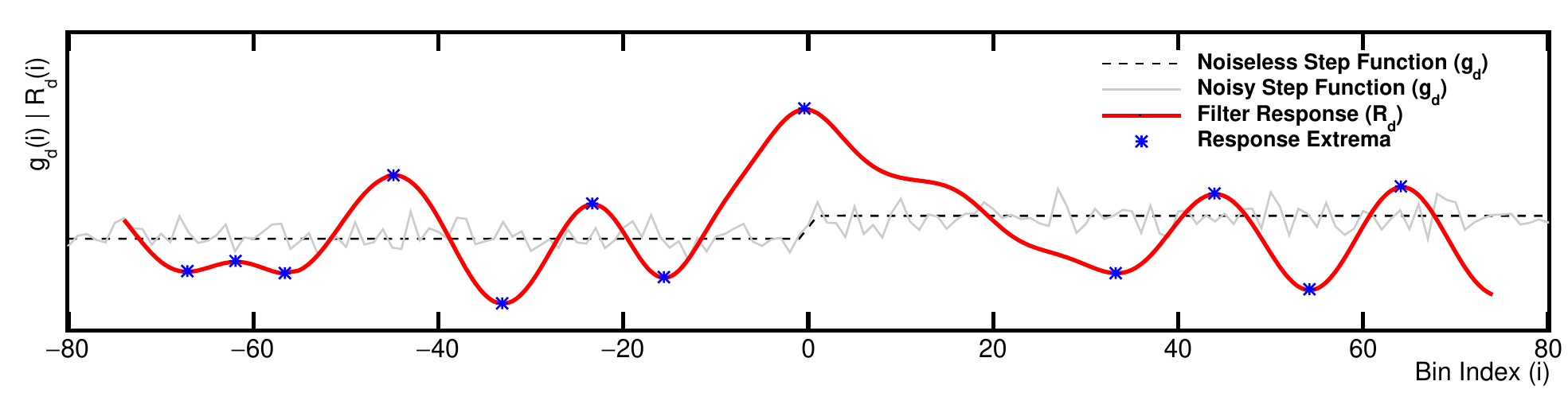}
	\caption{Example of the application of the edge finding algorithm to a noisy step function with a signal-to-noise ratio of 1.0. The test function to which the algorithm is applied is obtained by overlaying a white-noise of $\sigma=1$ to an ideal step function of amplitude $A=1$ at $i=0$. We processed the test function with a FDOG filter with $\sigma_f = 5$. Using our algorithm the maximum positive response which identifies the step position was detected at $x=-0.4$}
	\label{fig:EdgeDetectionExample}	
\end{figure}

Using those ideas, given a discrete physical distribution defined on a regular grid, e.g., a histogram with uniform binning, and a filter defined by its $2N +1$ coefficients $\bm{h_d(k)}$, we can define the position of an edge-like feature as the point where the input function's first derivative is maximum. Applying our algorithm we can compute this position calculating the position of the maximum of the response function $\bm{R_d(i)}$. A simple example of a noisy step function as test function $g_d$ and the response of a FDOG filter $R_d$ is shown in Fig.~\ref{fig:EdgeDetectionExample}. 

This technique is common in data processing. It is employed, e.g., for finding edges in digital images or for identifying a significant pulse in a noisy background when performing signal processing. In those contexts, the emphasis is to find an appropriate solution in large number of independent datasets, e.g., to identify the edges in millions of different images. In contrast, our study aims at defining the optimal filter to measure, as precisely as possible, the position of a specific feature, i.e. the edge, in a single distribution, taking into account the measurement errors of all samples. Furthermore, it is also crucial to determine a robust and reliable estimate for the uncertainty associated with the edge's position measurement. Therefore, the most important element of our study is the establishment and application of an optimization and uncertainty estimation procedure.


%

\subsection{Definition of Data Properties and Filter Types}\label{sec:FilterDefinitions}

\begin{figure}
	\centering
	\includegraphics[width=0.66\textwidth]{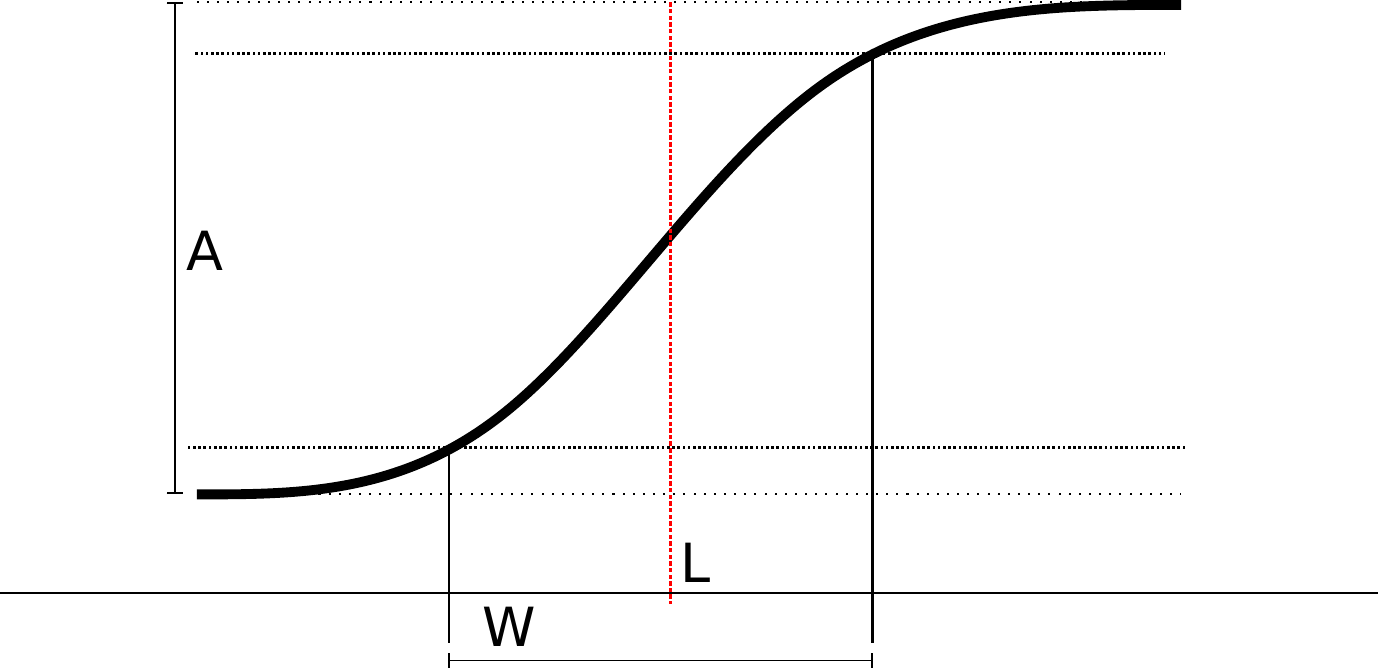}
	\caption{Definition of the important parameters used to characterize the edges. \textbf{L} represents the edge position, \textbf{A} the amplitude, \textbf{W} the width.}
	\label{fig:StepParametersExample}	
\end{figure}



The data distributions to be analysed for the presence of an edge can be characterized by the following properties, visualized in Fig.~\ref{fig:StepParametersExample}:
\begin{description}
       \item[Location of the edge (L):] The location $L$ of the edge is the main target observable. For an analytical function, this corresponds to the location of the maximum of the first derivative of the function.
       \item[Amplitude of the edge (A):] The amplitude $A$ of the edge is the total height of the edge and a measure of the signal strength.
       \item[Width of the edge (W):] The edge width $W$ which, in this work, we will define as the range in the function domain, or, in case of a histogram, the number of bins over which the distribution rises from $10$\% to $90$\% (or drops from $90$\% to $10$\%) of the edge amplitude $A$. For an ideal step function, $W=1$. When the step is defined by an analytical function, the width can also be calculated analytically and be defined as a real number.
       \item[Signal-to-Noise Ratio (SNR):] In case additional white noise is overlaid to the ideal function, we define the \emph{Signal-to-Noise Ratio} as the ratio between the maximum of the derivative \textbf{(S)} of the ideal function and the noise RMS.
\end{description}

In order to compare the performance of different filters, it is convenient to define a standard \emph{Scale Factor}, computed as the RMS of the \emph{Step Response}, that is the filter response to an ideal, noiseless step function with amplitude $A=1$, width $W=1$ and the edge located at $k=0$. \\
In the following, we will compare four different convolution filters:

\begin{description}
	\item[Difference of Boxes (DOB):] This filter is equivalent to the execution of a moving average in a window of size $2N +1$ prior to the derivation stage. The characterizing parameter of this filter is the size $\bm{N}$ of the averaging window and can be analytically written as:
	\begin{align}
	h_d(k) = 1 &\quad\text{for}\quad& -N \leq k < 0 \nonumber\\
	h_d(k) = 0 &\quad\text{for}\quad& k = 0 \nonumber\\
	h_d(k) = -1 &\quad\text{for}\quad& 0 < k \leq N
	\end{align}
	\item[First Derivative of a Gaussian (FDOG): ] This filter is equivalent to the execution of a Gaussian smoothing prior to the derivative. The filter is parametrised by the $\bm{\sigma}$ parameter of the Gaussian:
	\begin{equation}
	h_d(k) = -k \ e^{-\frac{k^2}{2\sigma^2}} \quad \text{for}\quad  -N \leq k  \leq N 
	\end{equation}
In case of the FDOG, the scale factor is equal to the width $\sigma$ of the Gaussian.
	\item[Deriche: ] This algorithm, described in \cite{Deriche}, is characterized by the following impulse response, parametrised by the exponential decay length $\bm{\alpha}$:
	\begin{equation}
	h_d(k) = -k \ e^{-\alpha |k|} \quad \text{for}\quad  -N \leq k  \leq N 
	\end{equation}
	\item[Shen: ] Similar to the previous one, this algorithm, described in \cite{Shen:1992:OLO:134450.134453}, is equivalent to an exponential smoothing and presents a discontinuity in the derivative at $k=0$:
	\begin{align}
	h_d(k) = e^{\alpha k} &\quad\text{for}\quad& -N \leq k < 0 \nonumber\\
	h_d(k) = 0 &\quad\text{for}\quad& k = 0 \nonumber\\
	h_d(k) = -e^{-\alpha k} &\quad\text{for}\quad& 0 < k \leq N
	\end{align}	
\end{description}

Additionally, we are going to compare the results of those filters to the numerical derivative, labeled in all figures as ``derivative'', equivalent, up to a constant, to a DOB filter with ${N=1}$. A detailed and extensive description of the characterization of these filters is given in~\cite{Caiazza:2018} and we will here summarize the most relevant findings. 


\begin{figure}
	\centering
	\includegraphics[width=\textwidth]{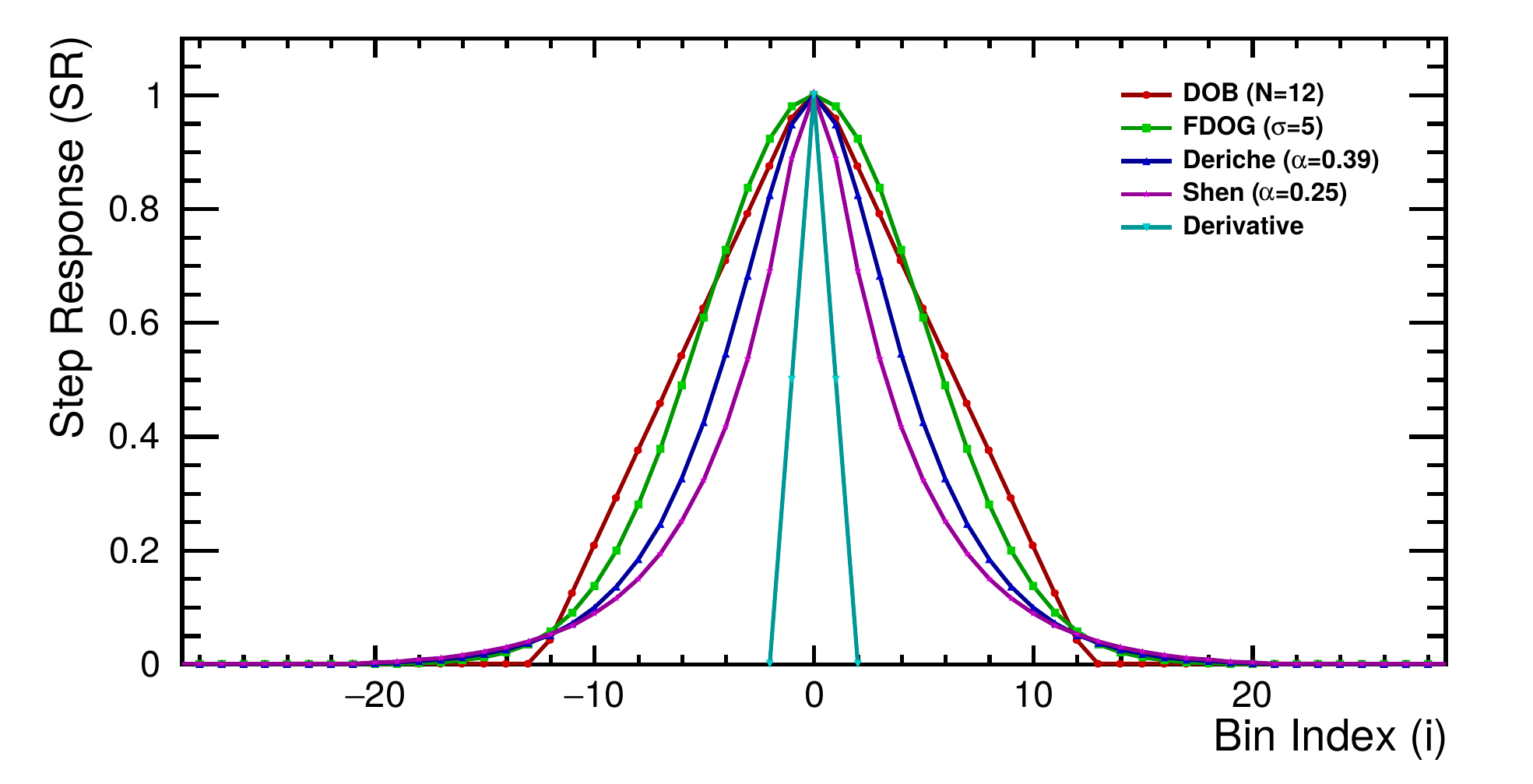}
	\caption{Normalized step response function (SR) of the 4 studied filters, as compared with the response of the numerical derivative. The characterizing parameter of each filter was chosen so that they will have a scale factor (RMS of the SR) of about 5 to make a comparison more meaningful.}
	\label{fig:FilterSRCompareNorm}	
\end{figure}

Those filters are graphically visualized through their \emph{Step Response} (SR) in Fig.~\ref{fig:FilterSRCompareNorm}. Thereby the characterizing parameters of all filters have been adjusted such that the scale factor is about 5 for each filter. The value of the scale factor has an impact on the amount of noise that the filter picks up during the convolution with the input distributions as well as on the precision of the edge localization. Qualitatively, we expected that when that value is too small, the filter response will be strongly influenced by the noise. In contrast, if it is too large, the correct edge will be found above the noise, but its position determination may be less precise and be influenced by the presence of other nearby features. We will quantify these aspects in the next sections.


\subsection{Optimisation criteria for a discrete function edge finder}\label{sec:FilterOptCriteria}
Using the tools introduced in the previous subsection we define two benchmark quantities: the filter efficiency and its localization power. Those are conceptually similar to those defined in~\cite{Canny:4767851} and we used them to characterize and optimize the filtering algorithm.
In all benchmarks presented in this section we use, as a test function, a Gaussian smoothed step of fixed amplitude with variable width and SNR. Using this function allows to define the width and the signal amplitude analytically as a function of the width parameter of the Gaussian smoothing $\bm{\Sigma}$ as: $\bm{W = 2 \sqrt{2 \ln 10} \Sigma}$ and $\bm{S = A / \sqrt{2 \pi} \Sigma}$. The Monte Carlo samples for the optimization criteria evaluation are generated using the aforementioned smooth step as a template, adding randomly generated white noise to each point in the function domain. The position of the grid points is randomized at every sampling to avoid any bias coming from the relative position of the function edge within a single subdivision of the grid.

\paragraph{Filter efficiency:}

\begin{figure}
	\centering
	\includegraphics[width=0.48\linewidth]{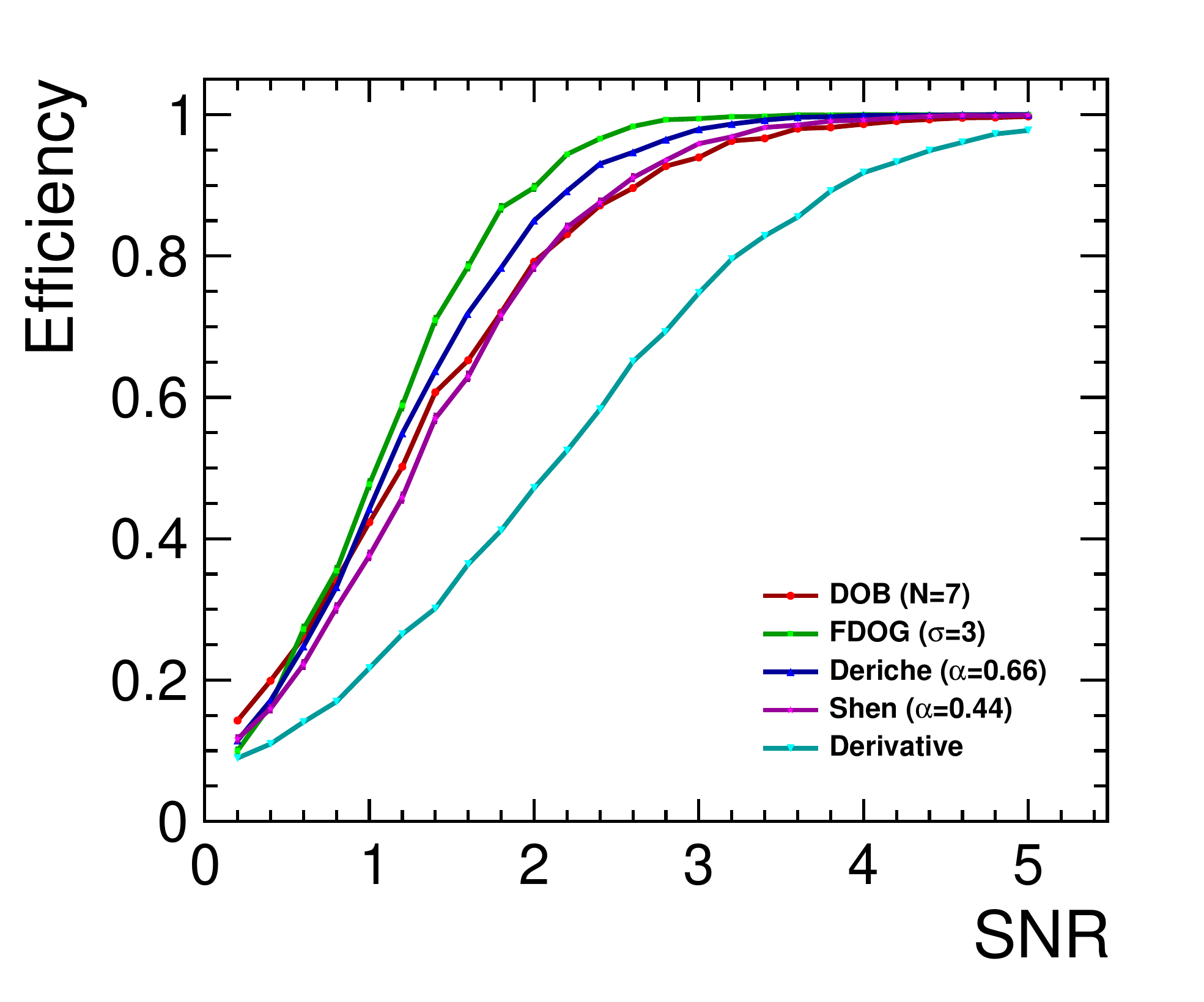}
	\includegraphics[width=0.48\linewidth]{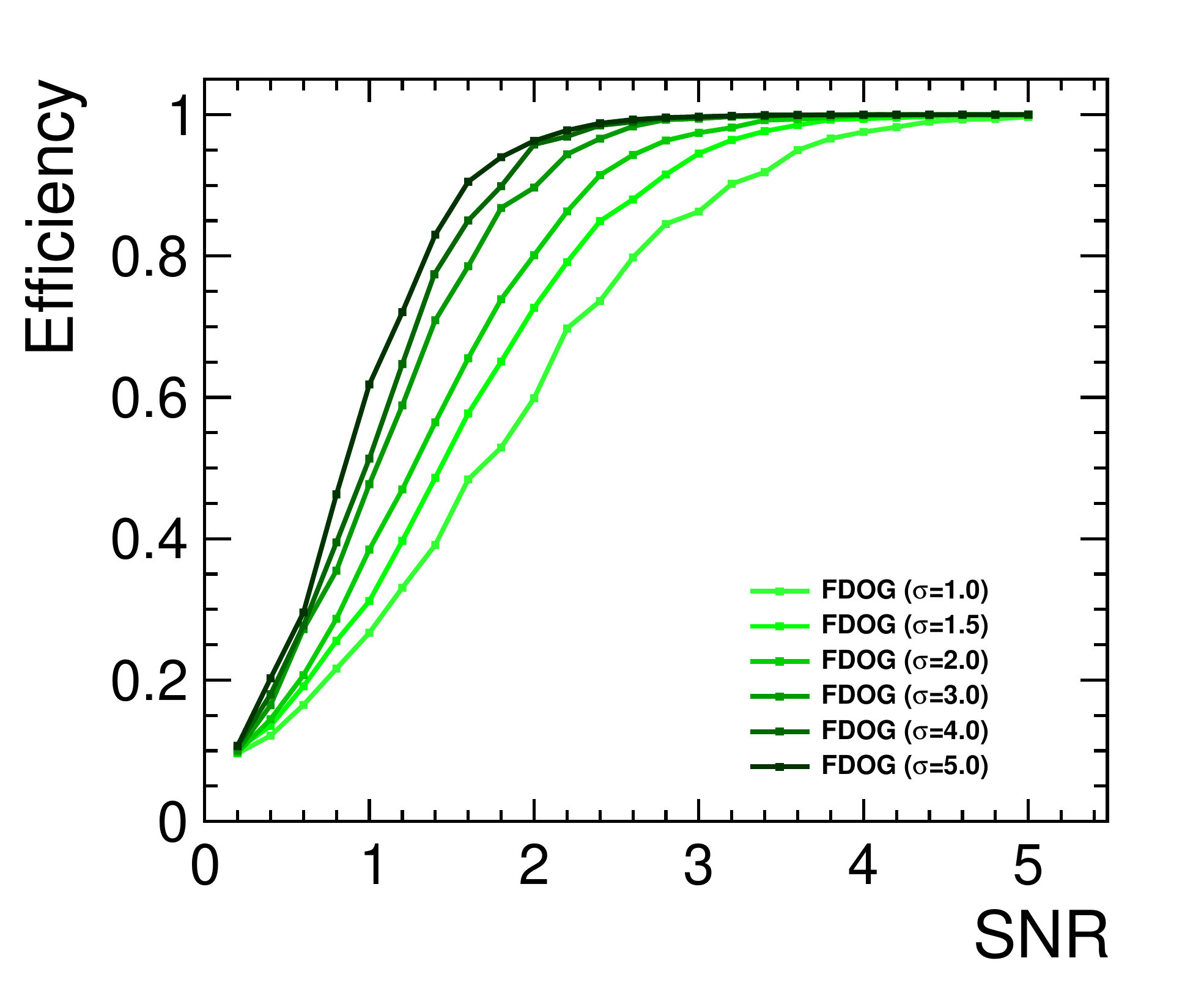}
	\caption[Comparison of the filter efficiency as a function of the SNR and of the filter size]{On the left, the plot shows the efficiency of four different filters, compared to that of the simple derivative, as function of the Signal-to-Noise-Ratio (SNR) of a noisy, smooth step function with a width $ W = 1$. Each point is obtained using 10000 samples of the same function with random white noise. The characterizing parameter of each filter was chosen so that they have a scale factor of about 3. On the right, a similar plot shows the efficiency of the same filter with changing scale factor.}
	\label{fig:FilterEffvsSNRS5}
\end{figure}

The filter efficiency, in the case of a rising edge, is defined as \emph{the probability that the local maximum of the filter's response closest to the true edge position of the input function is the global maximum}. In case of a falling edge it is sufficient to replace minimum with maximum in the previous definition. This probability can be calculated by processing the input function with different random noise samples an arbitrary number of times. This benchmark quantity was evaluated for the four studied filters and, for comparison, for the simple numerical derivative as a function of the SNR of the input distribution. 
As shown in fig.~\ref{fig:FilterEffvsSNRS5}, the efficiency of all filters increases with the SNR and increases monotonically with the filter scale factor at all SNR. At low SNR in particular, all studied filters perform significantly better than the simple derivative, which is more sensitive to noise than any of the filters. 

\paragraph{Localization power:}
The second parameter, the localization power, is particularly important for this work as it defines the systematic and statistical uncertainty that can be associated to the results we obtain. It comprises the \emph{localization bias} and the \emph{localization error}, which are defined as the mean $\bm{\hat{L}}$ and standard deviation $\bm{\sigma_L}$, respectively, of the detected edge positions obtained when applying the algorithm to many random variations of the same underlying function. In general, the localization is defined in unit of the grid granularity, e.g. the width of an histogram bin. The difference between $\hat{L}$ and the true edge position defines the systematic uncertainty of the filter output, while $\sigma_L$ corresponds to the statistical uncertainty, whose evaluation does not require any prior definition of a \emph{true} edge position. 

\begin{figure}
\centering
\includegraphics[width=0.48\linewidth]{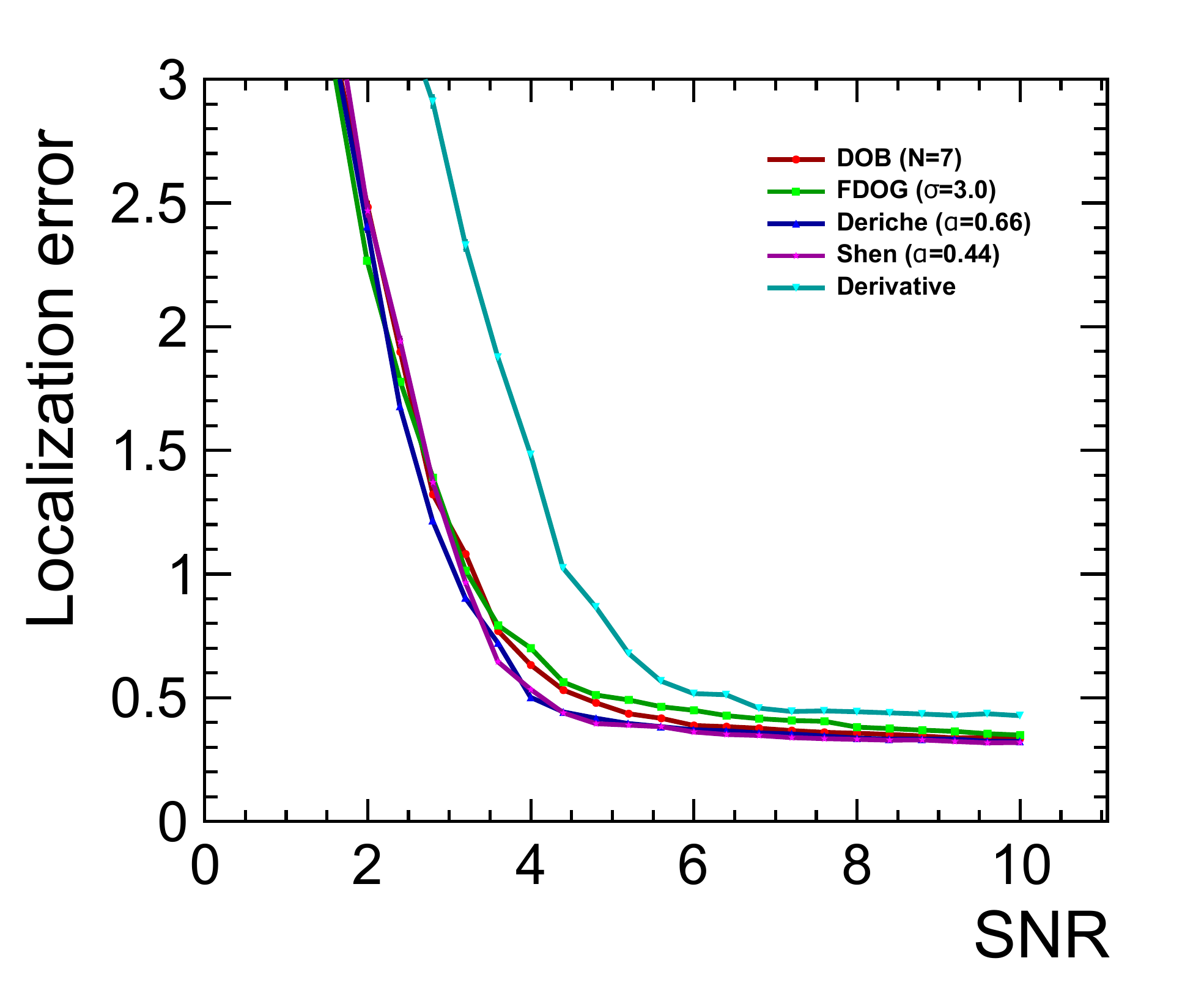}
\includegraphics[width=0.48\linewidth]{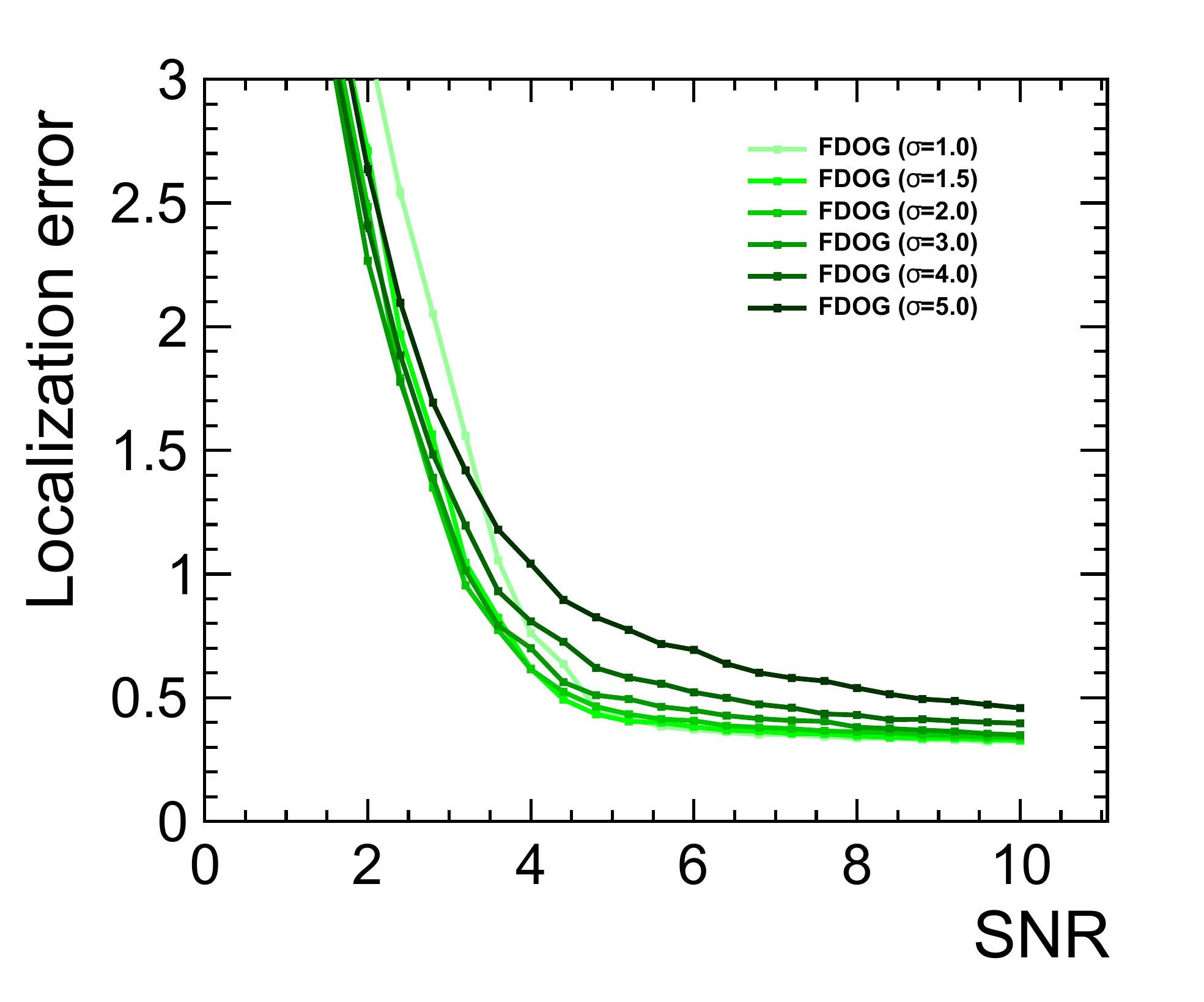}
\caption[Comparison of the localization error of different filters as a function of the SNR]{On the left, the plot shows the localization error of the four different filters, compared to that of the simple derivative, as function of the Signal-to-Noise-Ratio (SNR) of a noisy, smooth step function with a width $ W = 1$. Each point is obtained using 10000 samples of the same function with random white noise. The characterizing parameter of each filter was chosen so that they have a scale factor of about 3. Note that this scale factor was chosen for ease of comparison between all plots rather than to select the best possible filter parametrization. On the right, a similar plot shows the localization error of the same filter type when changing the scale factor.}
\label{fig:FilterSmoothLocvsSNR}
\end{figure}

\begin{figure}
	\centering
	\includegraphics[width=0.48\linewidth]{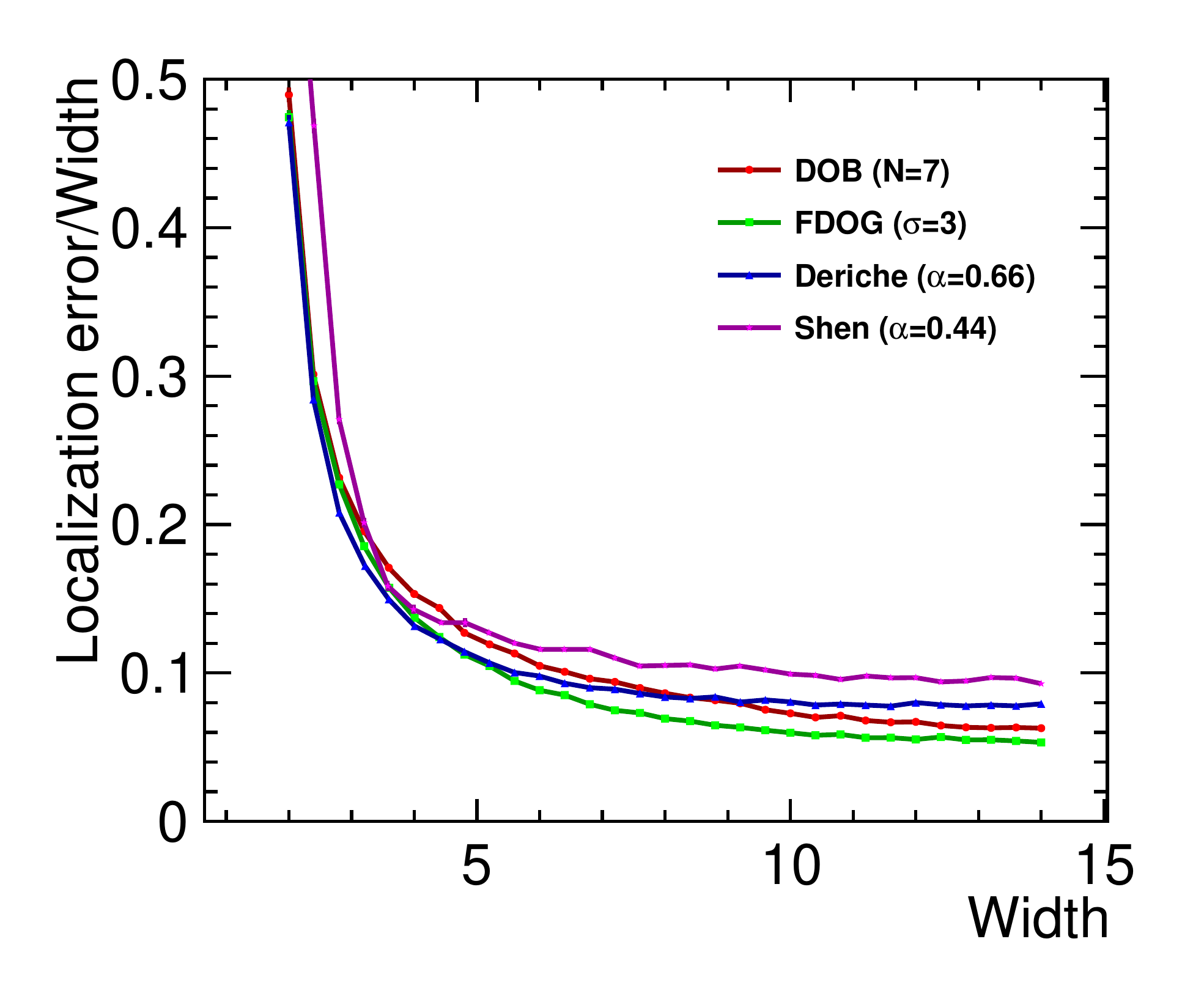}
	\includegraphics[width=0.48\linewidth]{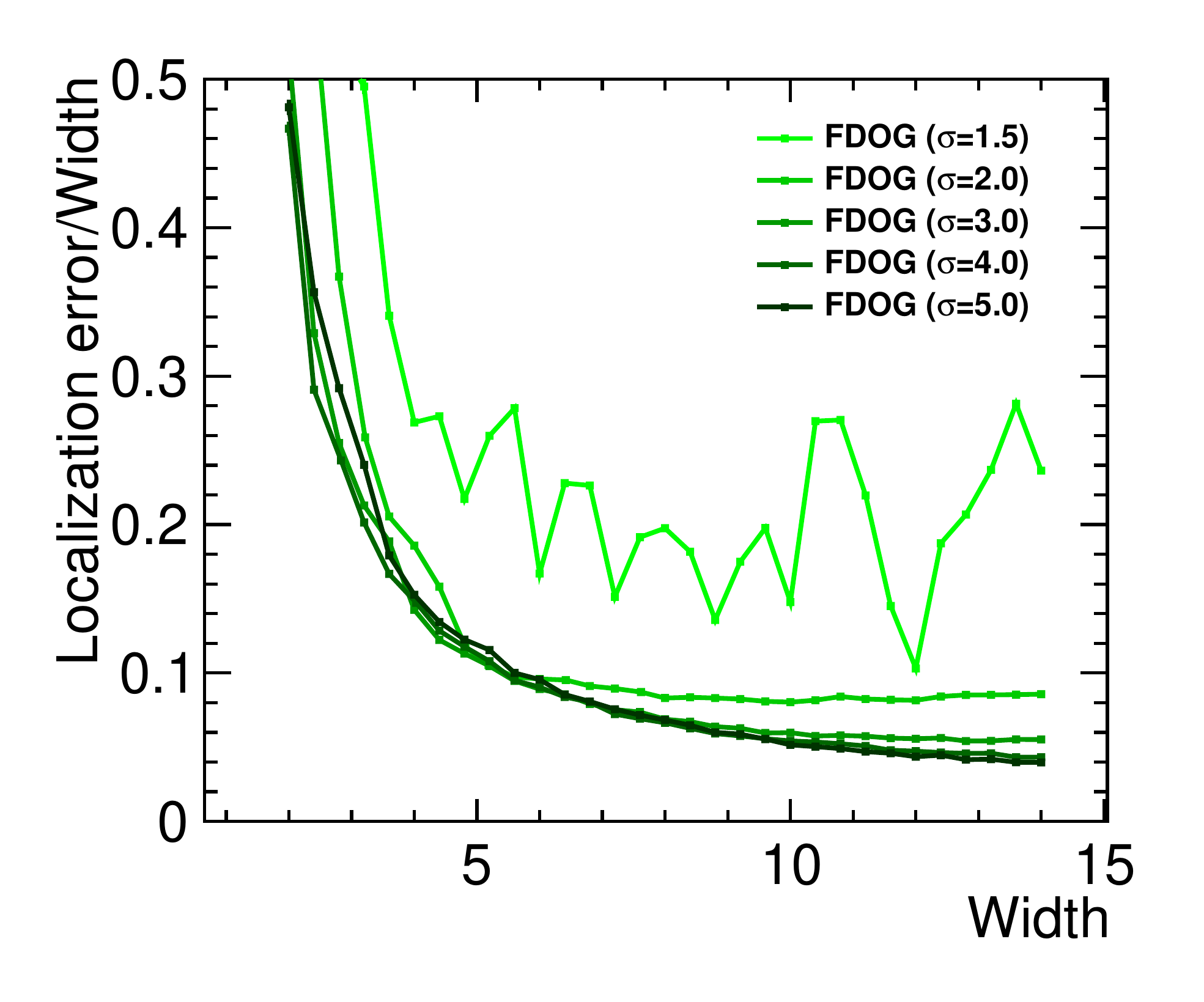}
	\caption[Comparison of the localization error of different filters as a function of the feature width]{On the left, the plot shows the normalized localization error of the four different filters, as a function of the width of a smooth step of varying width and Signal-to-Noise-Ratio (SNR), as described in the text. Each point is obtained using 10000 samples of the same function with random white noise. The characterizing parameter of each filter was chosen so that they have a scale factor of about 3. On the right the same is shown for a set of FDOG filters with different scale factors. The derivative is not shown in those plots as its localization is so poor to be out of scale.}
	\label{fig:FilterLocvsWidth}
\end{figure}

In our test cases, where the edge is defined on top of a constant baseline and the additive noise is uniform and doesn't cause an asymmetric distortion of the edge, the localization bias $\hat{L}$ is always compatible with zero in all conditions. This will change in the physics examples discussed in Sec.~\ref{sec:selectrons} and~\ref{sec:gauginos}, which include non-flat backgrounds and non-uniform bin errors. The localization error $\sigma_L$, on the other hand, has important variations depending on the SNR, on the edge width $W$ and the scale factor of the filter, which are illustrated in fig.~\ref{fig:FilterSmoothLocvsSNR} and~\ref{fig:FilterLocvsWidth}, respectively.

In relation to the noise level we observe that, at large enough SNR, all filters of similar scale factor are roughly equivalent, with a localization error decreasing asymptotically towards a value of about 30\% of the grid granularity. Moreover we can note that all are notably superior to a simple numerical derivative. We also note, looking at the right panel of fig.~\ref{fig:FilterSmoothLocvsSNR}, that there is an optimal filter configuration that provides the best localization and that this configuration can be chosen unambiguously using our optimization technique.

To transition to the discussion of the case studies, in which the function to analyse is obtained from a histogram with user-defined binning, the plots in fig.~\ref{fig:FilterLocvsWidth} show the ratio between the localization error and the step width, which we will call \emph{normalized localization}, rather than the simple localization error. Moreover the SNR overlaid on the function is proportional to $1/\sqrt{W}$, starting from an SNR of 3 for $W=2$. In this way we can emulate the case of an histogram where changing the bin width leads to a correlated change of the signal and background entries and thus the SNR. Here we notice again that all filters show a similar behaviour: the normalized localization improves rapidly with the increasing step width for small values of that quantity and tends asymptotically to a value that depends on the choice of the filter and on its scale factor.



\subsection{Filter choice and optimization}
\label{sec:edge-optimization}
With the results shown in the previous sections we can define a procedure to perform an accurate and optimal measurement of the position of edge-like features in discrete one-dimensional distributions like those obtained in particle physics experiments and a robust way to evaluate the statistical and systematic error to associate to such measurements.

Previously we have shown that the efficiency and localization power of all filters drop significantly when the SNR is reduced below a threshold that depends on the filter scale factor and type. Above that threshold larger filters tend to perform better but improvements quickly become negligible.

As we have shown in the previous section, the algorithm parameters can be optimized through a Monte-Carlo technique, by considering the input distribution a template from which an arbitrary number of samples can ge generated. It is important to note that this whole optimization can be performed without any knowledge of the specific shape of the feature to detect or of its theoretical position, if that could be calculated at all. In fact, this optimization does not require any prior knowledge of the underlying distribution.

When the grid resolution can be customized by the user, like in the case of an histogram generated from a large dataset, the first step is thus to optimize that resolution, e.g. the bin width of the input histogram, such that the average difference between the value at neighbouring positions in the step region is in the order of 25 entries, corresponding to a SNR of about 5, to be able to use the largest possible filter while keeping an optimal efficiency.

 
With the grid resolution fixed it is then possible to fine tune the algorithm and determine the localization errors for different filters of different scale factors, to find the combination which gives the smallest localization error. Unless the step to be localized is very narrow even after the adjustment of the grid resolution, the filter of choice is typically the FDOG, where the filter scale factor corresponds to the parameter $\sigma$.




\subsection{Calibration procedures for the detected edges}
To define the systematic accuracy of the edge position measurement we need to introduce some theoretical model of the underlying distribution that would define the \emph{true edge} position. By using such a model it is possible to use a toy Monte-Carlo procedure to calibrate the edge position measurement and to define its systematic accuracy. In general, this can be done by calculating the localization bias of the edge position measurement for each element in a set of model parameters influencing the true edge position so that a correlation between them can be derived. This correlation is model dependent and can be studied in each specific case to estimate the systematic error induced to the measurement for each set of model parameters. In the rest of this article, two physics benchmarks with very different challenges and different types of systematic uncertainties are studied, with the goal to validate the proposed method in actual physics analyses.

  \section{Study Cases}
\label{sec:studycases}

Many models of new physics predict additional particles which promptly decay into more stable particles, of which
some can be seen in the detector, while others may interact too weakly to leave any trace in the detector.
If all decay products can be detected, the mass of the new particle can simply be calculated as the invariant mass of all decay products. If there are, however, also ``invisible'' particles among the decay products, the detection of kinematic edges in the energy (momentum) or invariant mass spectra of the visible decay products is a well established technique to determine the mass of the mother particle.
This technique was widely used, for example, by the LEP experiments when performing supersymmetry 
searches~\cite{Abdallah:2003xe,ALEPH,OPAL,L3}.
%
Similar techniques are used in supersymmetry searches at LHC \cite{Hinchliffe:1996iu, Gjelsten:2004ki, Burns:2009zi, Matchev:2009iw, Galon:2011wh, Aaboud:2018ujj}.
Beyond supersymmetry, partially invisible decays are expected e.g.\ in most extensions of the Standard Model of particle physics which contain dark matter candidate particles.

In order to test and evaluate the performance of our proposed edge localisation method, we consider two different supersymmetry scenarios. Since at lepton colliders the well-known momenta of the colliding elementary particles allows a more complete reconstruction of the event kinematics, we especially study the production of supersymmetric particles at a future $e^+e^-$ collider. 
In both scenarios, the analysed reactions are of the generic type $e^+e^- \rightarrow \widetilde{Y} \bar{\widetilde{Y}}  \rightarrow \widetilde{\chi}^0_1 Y \widetilde{\chi}^0_1 \bar{Y}$, where the $\widetilde{\chi}^0_1$ is the lightest SUSY particle, assumed to be
stable and not detectable, $\widetilde{Y}$ is a generic SUSY particle,
and $Y$ its SM partner.

The kinematic edges comes about as follows:
in the rest-frame of a system with mass $M_0$ decaying to two particles
with masses $M_1$ and $M_2$, one has

\begin{align}
E_{1,(2)} = & 
\frac{ M^2_o + M^2_{1,(2)} - M^2_{2,(1)}}  { 2 M_0} \nonumber \\
p_1 = p_2 = p = & 
\frac { \sqrt { \lambda  \left ( M^2_0,M^2_1, M^2_2 \right ) } } { 2 M_0 } =  \frac { \sqrt { \lambda_{0,1,2} }}{ 2 M_0} \label{eq:epgeneric}
\end{align}
where the K\"all\'en function $\lambda$ is defined by
$\lambda(a,b,c) = a^2 + b^2 + c^2 - 2 ab -2 ac -2 bc$.

At production, $M_0 = E_{cms}$, and in the case of production
of a particle-antiparticle pair $\widetilde{Y} \bar{\widetilde{Y}}$
one has
$ M_1 = M_2 = M_{\widetilde{Y}} $, so that
$
E_1 = E_2 = E_{\widetilde{Y}} =
E_{cms}/2  $
~and
$
p = 
(E_{cms}/2) \sqrt { 1 - 
   \left ( M_{\widetilde{Y}}/ (E_{cms}/2) \right )^2 }.
$
 In the rest-frame of the decaying particle, 
one would identify $M_0 , M_1$~and $M_2$ in eq. \ref{eq:epgeneric} with
$M_{\widetilde{Y}}$,
$M_{\widetilde{\chi}^0_1}$~and $ M_Y$,
respectively.
The energy of $Y$ in the laboratory frame, $E_{Y~lab}$, is then given by

\begin{align}
E_{Y~lab} =& \gamma E_Y + \gamma\beta \cos{\theta} p_Y 
\end{align}

where $\gamma$ and  $\gamma\beta$ are the parameters of the Lorentz-transformation between
the laboratory frame and the rest frame of $\widetilde{Y}$.
The
highest and lowest possible energies are obtained for
$\theta = 0$ and $\theta = \pi$, respectively. Therefore,
the \textit{kinematic edges} are

\begin{align}
E_{Y~lab^{max}_{(min)}} =& \gamma E_Y  \begin{smallmatrix}+ \\ (-)\end{smallmatrix} \gamma\beta p_Y 
\end{align}

and are independent of the distribution of $\cos{\theta}$.

The relation between the position of the edges and 
the masses $M_{\widetilde{Y}}$~and
$M_{\widetilde{\chi}^0_1}$,
can be found by noting that
the Lorentz factors of the transformation from the laboratory frame to
the rest-frame of the produced particle are
$
\gamma = 
E/M =   E_{CMS}/( 2 M_{\widetilde{Y}} )   $
~and
$
\gamma \beta = 
p/M $~=$  E_{cms}/(2  M_{\widetilde{Y}} ) $~$\sqrt { 1 - 
   \left ( M_{\widetilde{Y}}/( E_{CMS}/2) \right )^2 },
$ 
~so that
the edges of the $E_{Y~lab}$ spectrum are  given by

\begin{align}
E_{Y~lab^{max}_{(min)}} 
    =&
 \frac{E_{CMS}} { 4 } 
\left (  \frac{ M^2_{\widetilde{Y}} + M^2_Y - M^2_{\widetilde{\chi}^0_1} }  {M^2_{\widetilde{Y}} } 
  \begin{smallmatrix}+ \\ (-)\end{smallmatrix} \sqrt{ 1 - \left ( \frac{M_{\widetilde{Y}}} { E_{CMS}/2} \right )^2 }
\frac{\sqrt { \lambda_{\widetilde{Y},Y,\widetilde{\chi}^0_1} }  } {M^2_{\widetilde{Y}} } 
\right ) \label{eq:minmaxanym}
\end{align}

In the study-case discussed in Sec.~\ref{sec:selectrons}, one might set $M_Y = 0$ , in which case
$E_Y = p_Y = 
(M^2_{\widetilde{Y}} - M^2_{\widetilde{\chi}^0_1} )/ ( 2 M_{\widetilde{Y}}), 
$
so that eq. \ref{eq:minmaxanym}~simplifies to

\begin{align}
E_{Y~lab^{max}_{(min)}} =&
 \frac{E_{CMS}} { 4  } \left ( 1 - \frac{M^2_{\widetilde{\chi}^0_1}}{M^2_{\widetilde{Y}} } \right ) 
    \left ( 1 
   \begin{smallmatrix}+ \\ (-)\end{smallmatrix} \sqrt { 1 - \left ( \frac{M_{\widetilde{Y}}} { E_{CMS}/2} \right )^2 } \right ) \label{eq:minmaxm0}
\end{align}

In eq. \ref{eq:minmaxm0}, it is easy to see that by calculating the sum of the
upper and lower edge of the spectrum ($\Sigma$), $M^2_{\widetilde{\chi}^0_1}$ can be expressed
in $M^2_{\widetilde{Y}}$, which can then be used to eliminate the former in the
expression for the difference between the edges ($\Delta$).
From the obtained expression,
 $M^2_{\widetilde{Y}}$ is obtained,
and by back-substitution also  $M^2_{\widetilde{\chi}^0_1}$ is obtained.

\begin{align}
M_{\widetilde{Y}}    = &     
   \frac{E_{CMS}}{2}  \frac{ \sqrt{ \Sigma^2 -\Delta^2 }} { \Sigma } \nonumber \\
 M_{\widetilde{\chi}^0_1}= &  \frac{E_{CMS}}{2} \sqrt{\frac{ E_{CMS} - 2  \Sigma }{E_{CMS}}}  \frac{ \sqrt{ \Sigma^2 -\Delta^2 }} { \Sigma }\label{eq:massesfredesm0}
\end{align}

Similar, albeit more complicated, expressions can also be derived
from eq. \ref{eq:minmaxanym}.

The two example scenarios, which will be introduced in more detail in the following subsections, were studied in the context of the International Linear Collider (ILC)~\cite{Bambade:2019fyw, Behnke:2013xla, Adolphsen:2013jya, Adolphsen:2013kya} and the International Large Detector (ILD)~\cite{Abe:2010aa,Behnke:2013lya}. 

The ILC is a planned electron-positron linear collider. The construction of its first stage with \newline $\sqrt{s}=$\unit[250]{GeV} is under political consideration in Japan. In this article we consider an ILC centre-of-mass energy of $\sqrt{s}=$\unit[500]{GeV}, an integrated luminosity of $\int\mathcal{L}dt$ = \unit[500]{fb}$^{-1}$ and a beam polarisation configuration of 80\% right-polarised electrons and 30\% left-polarised positrons, i.e., $\mathcal{P}$($e^{-}$, $e^{+}$) = (+80\%, -30\%). This corresponds to only small fraction of the total data expected from the full run plan of the ILC~\cite{Fujii:2017vwa, Barklow:2015tja}.
All Monte-Carlo events used in the two case studies have been generated with \texttt{Whizard} 1.95~\cite{Kilian:2007gr}, with \texttt{PYTHIA~6.4} \cite{Sjostrand:2006za} used for fragmentation and hadronization, and  taking into account the beam energy spectrum, initial and final state radiation, and the effects of beamsstrahlung. 


ILD is one of the two detector concepts proposed for the ILC, which were specifically designed according to the Particle Flow principles. ILD will be equipped with a time projection chamber acting as its main tracker and a very granular calorimetric system, with a fine longitudinal and transverse segmentation, optimized for particle flow reconstruction. The ILD design is implemented in a detailed \texttt{Geant4-}based simulation~\cite{MoradeFreitas:2002kj, AGOSTINELLI2003250}, and the simulated detector response is reconstructed by the algorithms implemented in the \texttt{Marlin} framework~\cite{Gaede:2006pj}, including tracking as well as the particle flow algorithm \texttt{PandoraPFA}~\cite{Marshall:2015rfa, Marshall:2012ry, Thomson:2009rp}.

\subsection{The STC4 Scenario}
\label{sec:STC4-intro}

The $\tilde{\tau}$-coannihilation scenario (STC) has been defined and motivated in \cite{PhysRevD.88.055004} in the theoretical framework of the phenomenological minimal supersymmetric Standard Model (pMSSM) \cite{Baer:1993ae, DJOUADI2007426}. It is characterised by a light slepton sector and heavy squarks that may have escaped detection at the LHC. The $\tilde{\tau}$ is the next-to-lightest supersymmetric particle with a mass approximately \unit[10]{GeV} larger than that of the $\tilde{\chi}_{1}^{0}$, i.e., the stable LSP (\emph{M}$_{\tilde{\chi}_{1}^{0}}$ = \unit[95.6]{GeV}). 

Several different STC model points have been defined \cite{Berggren:2015qua}, which differ mainly in the value assumed for the right-handed third generation squark mass parameter at \unit[1]{TeV}. In case of the STC4 point, considered in our evaluation, the value of this parameter is assumed to be \unit[400]{GeV}. The part of the spectrum relevant to an $e^+e^-$ collider with $\sqrt{s}=$\unit[500]{GeV} depends only weakly on this parameter and is for all practical purposes the same for all model points of the STC series.

Our study concentrates on selectron pair production at the ILC and their subsequent, almost exclusive decay: $e^{+}e^{-} \rightarrow \tilde{e} \tilde{e} \rightarrow e^{\pm}\tilde{\chi}_{1}^{0} e^{\mp}\tilde{\chi}_{1}^{0}$. This takes advantage of the relatively large cross-section that characterises selectron pair production in the STC4 scenario, due to the t-channel neutralino exchange: $\mathcal{O}$($10^{2}$ {fb}) for the $\tilde{e}_{R}\tilde{e}_{R}$ case and $\mathcal{O}$(10 {fb}) for the $\tilde{e}_{L}\tilde{e}_{L}$ case. Furthermore, the event selection also benefits from the higher efficiency to identify electrons in contrast to $\tau$ decays. Lastly, if \emph{M}$_{\tilde{\chi}_{1}^{0}}$ is extracted from the selectron momentum spectrum the $\tilde{\tau}$ mass can also be determined, using the same data set, from the endpoint of the spectrum of the $\tau$ decay products. 

Based on the assumed STC4 parameter values, the mass spectrum calculator \texttt{SPheno} \cite{Porod:2003um, Porod:2011nf} predicts a value of \emph{M}$_{\tilde{e}_{L}}$ = \unit[213.1]{GeV} for the left-handed selectron mass and \emph{M}$_{\tilde{e}_{R}}$ = \unit[126.2]{GeV} for its right-handed counterpart. The CMS experiment~\cite{Sirunyan:2018nwe, Aaboud:2018jiw} has already excluded  left-handed selectrons in the mass range assumed by the STC4 model. However, their right-handed partners have not yet been excluded. They are only $\sim$\unit[30]{GeV} heavier than the LSP and, due to this  small mass difference, the amount of missing energy in the LHC events would not be sufficient to allow a clear separation from the Standard Model background. 

While theoretically the isotropic decay of a scalar particle in its rest frame is expected to lead to a pure box-like distribution of the lepton momentum in the lab frame, the initial state radiation and the interactions of the final-state electrons in the detector distort the shape of the decay product momentum distribution so that it cannot be reasonably described by any specific analytic function calculated from first principles. Therefore this scenario is an excellent benchmark for our edge detection algorithm. 
Moreover, the same simulated data was used to estimate the same quantity using a different technique~\cite{Berggren:2015qua}, which can serve as a direct reference to estimate the performance of our method. In that publication, the method relied on the fact that, for an undistorted rectangular distribution without background, estimators surpassing the CRB (the  Cram\'er-Rao Bound) for the minimal variance exist for the high and low edges\footnote{The reason that the CRB can be surpassed in the case of the rectangular distribution is that this distribution does not fulfil one of the conditions of applicability of the Cram\'er-Rao bound, namely that the support of the density does not depend on the fitted parameter}. They are simply the smallest and largest observations in the sample, respectively. These statistics are only asymptotically unbiased, but the finite sample-size bias is analytically known. The variance of the estimators decreases as the reciprocal of the {\emph square} of the sample size. To mitigate the distortions and the presence of background, the approach was to divide the sample in sub-samples, and estimate the edges, not by the highest and lowest observations, but to first exclude a fraction of the extreme values. The edge estimates were then taken as the average of the truncated sub-sample extremes. The two parameters - sub-sample size and truncation factor - were optimised to give the smallest variance using a toy Monte-Carlo procedure. The toy MC was also used to calibrate the estimators in order to remove the bias\footnote{The bias due to a truncation of a pure rectangular distribution is analytically calculable, but not the effects of background and distortions.}. It was found that  \emph{M}$_{\tilde{e}_{R}}$ could be determined to \unit[210]{MeV} and \emph{M}$_{\tilde{\chi}_{1}^{0}}$ to \unit[160]{MeV} based on an integrated luminosity of \unit[500]{fb}$^{-1}$ with $\mathcal{P}$($e^{-}$, $e^{+}$) = (+80\%, -30\%) at a center-of-mass energy of \unit[500]{GeV}.

\subsection{The Point 5 Scenario}
\label{sec:Point5-intro}

The supersymmetry scenario denoted as "\emph{Point 5}" was first defined \cite{Battaglia:2006bv} in the context of the minimal Supergravity model \cite{Chamseddine:1982jx, Barbieri:1982eh}, mSUGRA. It was originally considered for benchmarking the Particle Flow performance of the ILD detector concept. 

In the Point 5 model, the first chargino ($\tilde{\chi}_{1}^{\pm}$) and the second neutralino ($\tilde{\chi}_{2}^{0}$) are almost mass degenerate: \emph{M}$_{\tilde{\chi}_{1}^{\pm}}$ = \unit[216.5]{GeV} and \emph{M}$_{\tilde{\chi}_{2}^{0}}$ = \unit[216.7]{GeV}. They could be produced in pairs at the ILC with an operating centre-of-mass energy of $\sqrt{s}$ = \unit[500]{GeV}. The production cross-section is of the order of $10^{2}$\,fb in the chargino case and of the order of $10$\,fb in the neutralino case. 

Both gauginos decay with branching ratios larger than 96\% to the corresponding gauge boson and the first neutralino which is the stable LSP:
\begin{eqnarray}
e^{+}e^{-} &\rightarrow &\tilde{\chi}_{1}^{+}\tilde{\chi}_{1}^{-} \rightarrow W^{+}\tilde{\chi}_{1}^{0}W^{-}\tilde{\chi}_{1}^{0} \nonumber \\
e^{+}e^{-} &\rightarrow& \tilde{\chi}_{2}^{0}\tilde{\chi}_{2}^{0} \rightarrow Z^{0}\tilde{\chi}_{1}^{0}Z^{0}\tilde{\chi}_{1}^{0} 
\label{eq:point5Reactions}
\end{eqnarray}
Their mass difference with respect to the LSP mass (\emph{M}$_{\tilde{\chi}_{1}^{0}}$ = \unit[115.7]{GeV}) is approximately \unit[100]{GeV}, i.e. large enough that the \emph{W} and \emph{Z} bosons can be produced on-shell. 

Our study aims to reconstruct and determine the masses of all three gauginos: $\tilde{\chi}_{1}^{\pm}$, $\tilde{\chi}_{2}^{0}$ and $\tilde{\chi}_{1}^{0}$. Consequently, each one of the two reactions presented above acts as background for the other. The separation of the chargino and neutralino candidate events essentially relies on identifying and distinguishing \emph{W}$^{+}$\emph{W}$^{-}$ and \emph{Z}\emph{Z} decays. Only the hadronic decays of the gauge bosons are here considered as signals. This was a deliberate choice, posing a significant challenge both for the Particle Flow reconstruction and for the proposed edge detection method. In practice, the (semi-)leptonic modes would of course be included, although their contribution is statistically limited.

The relevant observables for determining the gaugino masses are the di-jet energy spectra, where each di-jet corresponds to a gauge boson. Our proposed edge localisation method must handle the smearing effects on the edges of the di-jet energy distributions produced by the beam energy spectrum. In addition, it must also deal with the contributions coming from the intrinsic gauge bosons' widths: $\Gamma_{W} \simeq$ \unit[2.1]{GeV} and $\Gamma_{Z} \simeq = \unit[2.5]{GeV}$ \cite{Tanabashi:2018oca}. The additional smearing effect due to the finite jet energy resolution is rather small compared to these two.

In the original Point 5 study \cite{Suehara:2009bj}, the positions of the edges were determined by fitting the di-jet energy  with a function in which the edge positions were free parameters. In order to model the signal, a second order polynomial, describing the  part of the spectra between the edges, was used. This was convoluted with a Voigt function in order to model both the effect of the width of the gauge bosons and of the uncertainties related to the measurement of jet energies. The Standard Model background contributions were assumed to be well understood by the time of ILC operation, and thus they were described independently by another heuristic function $f_{SM}$ which was fixed in the edge position fit:   

\begin{equation}
 f_{total} = f_{SM} + \int_{t_{0}}^{t_{1}} (a_{sig}\cdot t^{2} + b_{sig}\cdot t + c_{sig})Voigt(t-x, \sigma, \Gamma)\mathit{dt}
\label{eq:globalFit}
\end{equation}
The convolution limits ($t_{0}$ and $t_{1}$) represent the edge positions and they are free parameters of the function, together with $a_{sig}$, $b_{sig}$, $c_{sig}$, $\sigma$, and $\Gamma$.

In order to evaluate the robustness of the fitting method, a toy Monte Carlo study was performed. For this purpose, 10$^{4}$ di-jet energy distributions (for each of the two gaugino candidates' samples) were produced as random variations of the histograms obtained in the full detector simulation-based analysis. The fitting function given by Eq.~\ref{eq:globalFit} was applied for each one of the newly generated distributions. From 10$^{4}$ performed fits, 8808 converged for the chargino case and 8282 for the neutralino case.  
		
\begin{figure}[h!]
		\begin{centering}
		\includegraphics[width=0.5\textwidth]{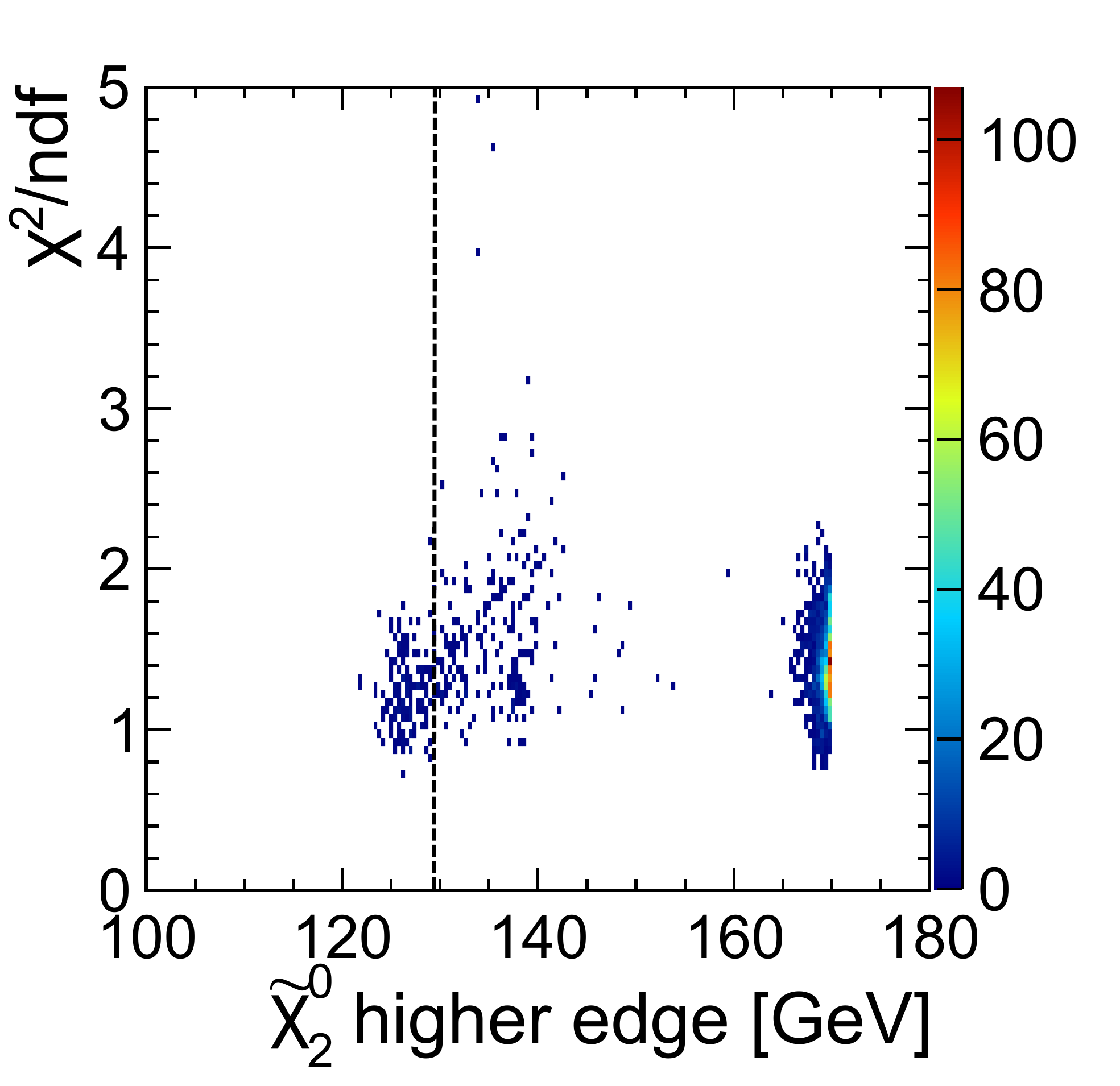}
		\caption{Results of the toy Monte Carlo study performed to evaluate the consistency of the fitting method for kinematic edge extraction. The number of converged fits is illustrated by the colour code. The black, dashed vertical line shows the location of the model calculated edge.}
		\label{fig:fitStability}
	\end{centering}
\end{figure}

In Fig.~\ref{fig:fitStability}, the $\chi^{2}/$ndf ratio, as a measure of the fit performance, is plotted with respect to the determined edge position for the neutralino upper edge as an example. The information concerning the number of fits associated with a specific $\chi^{2}/$ndf - edge position configuration can be read from the colour code. 

The instability of the fitting method is indicated by the wide range of obtained edge values. It can be seen from Fig.~\ref{fig:fitStability} that the obtained edge values vary over a range of approximately \unit[40]{GeV} despite the apparently good quality of the fit ($\chi^{2}/$ndf < 1.5). This effect is significantly larger than any expected shift in the edge position that could occur from the random generation of the distribution. Furthermore, the majority of fits falsely determined that the higher edge positions coincide with the histrograms' end points at \unit[170]{GeV}.

Thus, as a result of the toy Monte Carlo study it could be assessed that the fitting method is highly sensitive to small fluctuations in the di-jet energy spectra. This strongly illustrates the need to develop a more robust edge detection method. We will show in the following sections that the FIR filter approach is significantly more stable and accurate in measuring the edge positions.

  \section{Measurement of the $\tilde{e}$ and $\tilde{\chi}^0_1$ masses at the ILC }
\label{sec:selectrons}

\newcommand\fnurl[2]{%
	\href{#2}{#1}\footnote{\url{#2}}%
}

The first study case is the measurement of the masses of the  $\tilde{e}_{R}$ and the $\tilde{\chi}^0_1$, respectively ${M_{\tilde{e}_{R}}}$, ${M_{\tilde{\chi}^0_1}}$, from the position of the kinematic end-points of the momentum distribution of the lepton produced in the final state. For this analysis, the signal and the full SM background were processed by the fast simulation program SGV~\cite{Berggren:2012ar} which models the response of the tracking system of the ILD detector from first principles, complemented by a parametrised response of the particle-flow calorimeters. Those data are finally processed by an electron identification algorithm, described in \cite{Caiazza:2018}. This tool was made specifically to operate on the SGV output to identify isolated electrons and the photons they may have radiated before entering the calorimeters.

In the process we are analysing, the mass of the two visible leptons in the final state is negligible compared to the masses of the other particles involved. Therefore we can use eq.~\ref{eq:minmaxm0} to estimate the energy distribution of those electrons, which would ideally be uniform between two sharp, rectangular edges whose positions we can calculate using eq.~\ref{eq:massesfredesm0}. In the STC4 scenario, with $M_{\tilde{\chi}_{1}^{0}}$ = \unit[95.6]{GeV} and $M_{\tilde{e}_{R}}$ = \unit[126.2]{GeV} and a centre-of-mass energy of \unit[500]{GeV}, the ideal position of the kinematics edges of the $e^+ e^-\rightarrow \bar{\tilde{e}}_R \tilde{e}_R \rightarrow e^+ e^- + 2 \tilde{\chi}^0_1$ are $E_e^{max}$ = \unit[99.362]{GeV} and $E_e^{min}$ = \unit[7.298]{GeV}. 

Compared to these ideal values, when considering the realistic distribution of the centre-of-mass energies, initial-state radiation and detector uncertainties, we expect the actual positions of the kinematic end-points to be shifted in a way that cannot be predicted analytically, therefore making this case an ideal test to measure the performances of our method.

\begin{figure}
	\centering
	\includegraphics[width=0.9\linewidth]{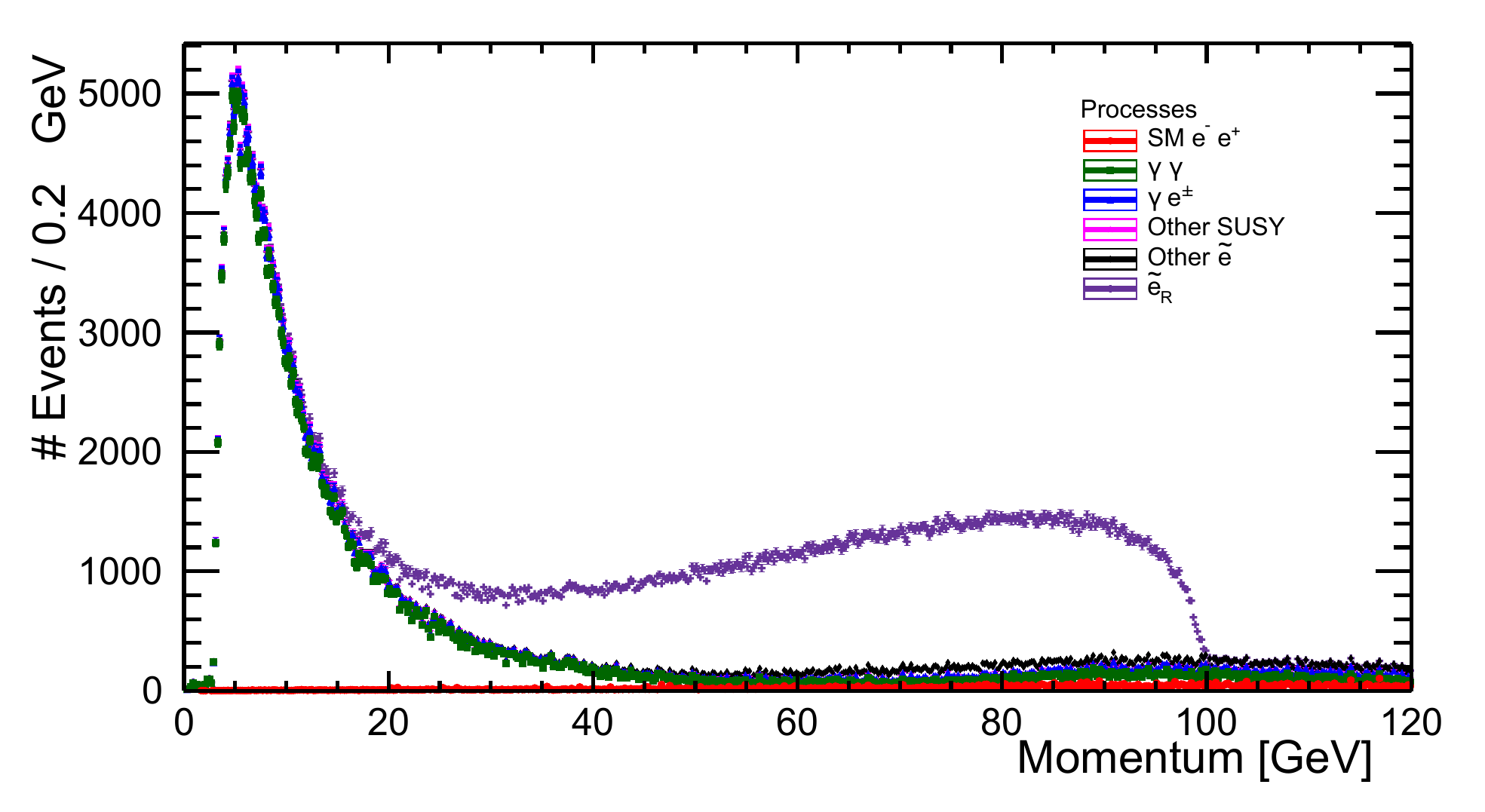}
	\caption[Momentum distribution of the higher-momentum electron after the high-momentum edge selection]{Momentum distribution of the higher-momentum electron after application of the high-momentum edge selection. This distribution is used as a template to calculate the position of the high-momentum edge and its uncertainty. }
	\label{fig:PMax_FinalSpectrum}
\end{figure}

\begin{figure}
	\centering
	\includegraphics[width=0.9\linewidth]{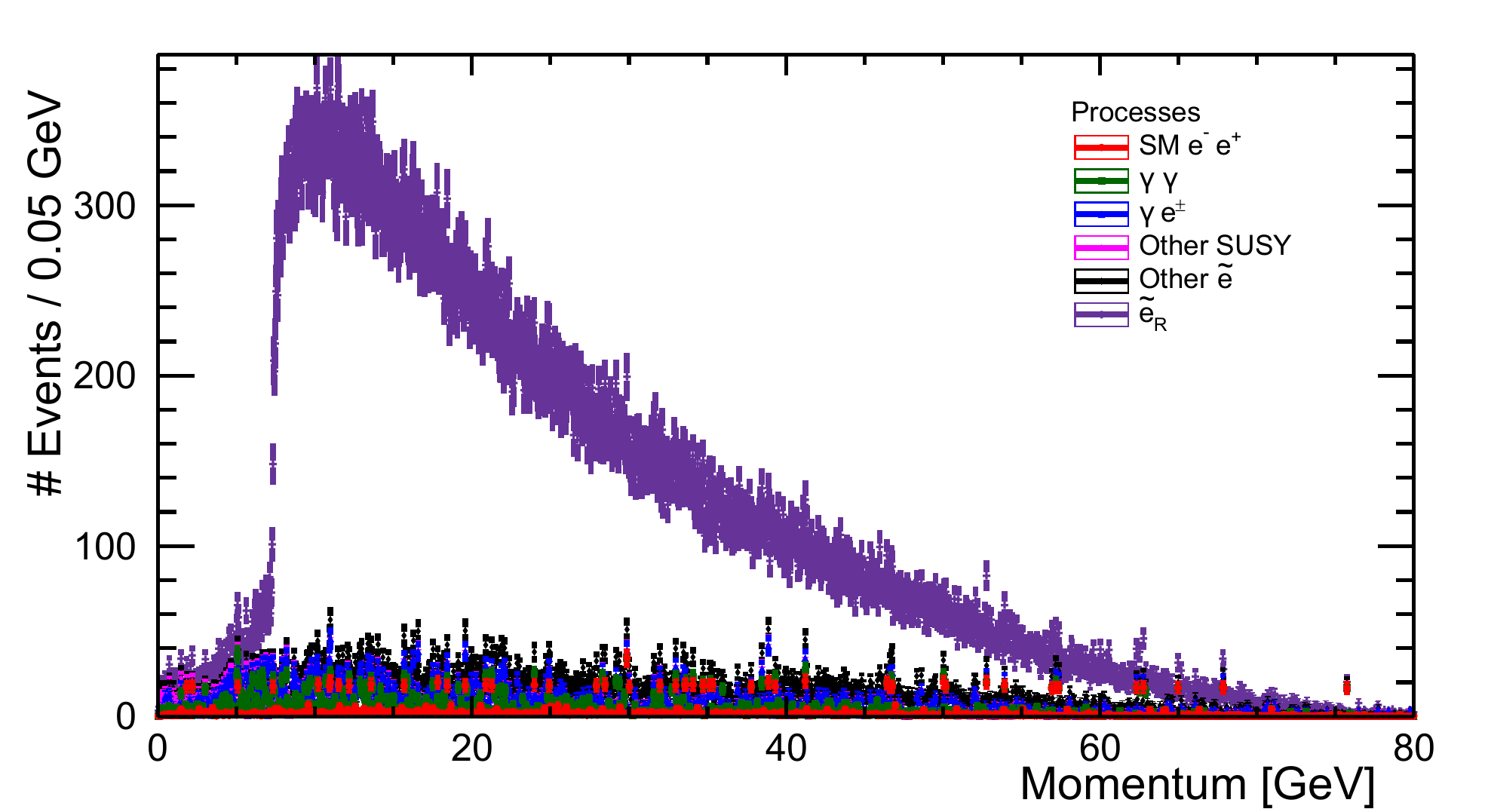}
	\caption[Momentum distribution of the lower-momentum electron after the low-momentum edge selection]{Momentum distribution of the lower-momentum electron after application of the low-momentum edge selection described in the text. This distribution is used as a template to calculate the position of the low-momentum edge and its uncertainty.}
	\label{fig:PMin_FinalSpectrum}
\end{figure}

Because the magnitude and sources of the distortions can depend on the energy of the final-state particle and because the backgrounds are significantly different at the location of the two end-points, we analyse them separately. In both cases the selection is based on cuts which identify events with negligible hadronic activity, large amounts of missing energy and exactly two isolated electrons of opposite charge. To reduce the $\gamma \gamma$ background, which dominates at low electron momenta, a cut on the acolinearity and acoplanarity of the two identified electrons with the higher momentum is applied. The momentum distribution of those two electrons is finally used to identify the high- and low-momentum end-points. Those distributions, after the event selection, are shown in Figs.~\ref{fig:PMax_FinalSpectrum} and~\ref{fig:PMin_FinalSpectrum}. Taking the whole data set into account, for the high-momentum selection we achieved a signal purity of 56.5\% with an efficiency of 82\% while for the low-momentum selection we achieved a signal purity of 87.6\% with an efficiency of 30\%. Full details of the selection can be found in \cite{Caiazza:2018}. 

\subsection{Edge position measurement and calibration}

\begin{figure}
	\centering
	\begin{subfigure}[b]{0.49\linewidth}
		\centering
		\includegraphics[width=\textwidth]{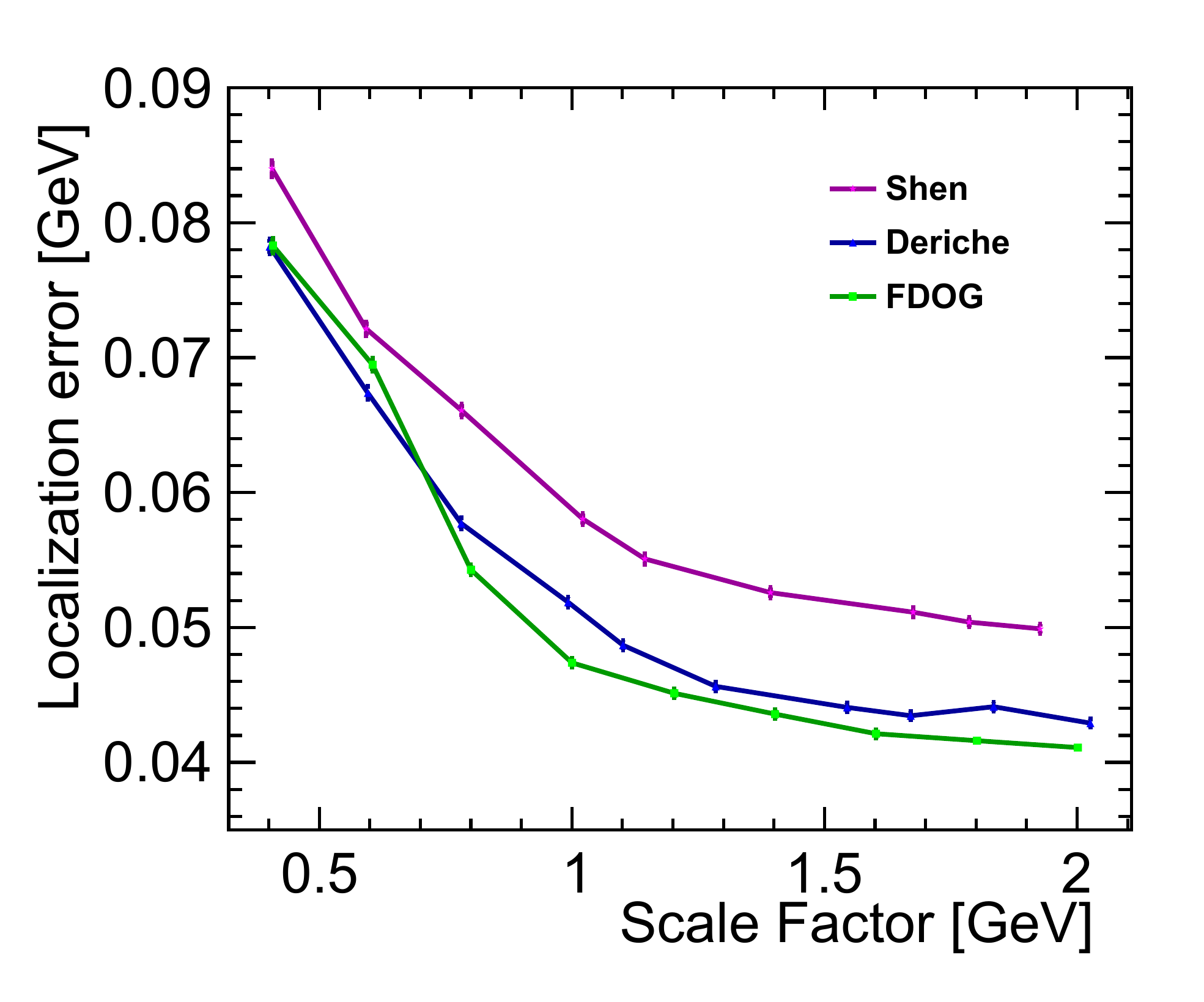}
		\caption[]{High-momentum edge optimization}
		\label{fig:PMax_LocOptimization}
	\end{subfigure}
\hfill
	\begin{subfigure}[b]{0.49\linewidth}
		\centering
		\includegraphics[width=\textwidth]{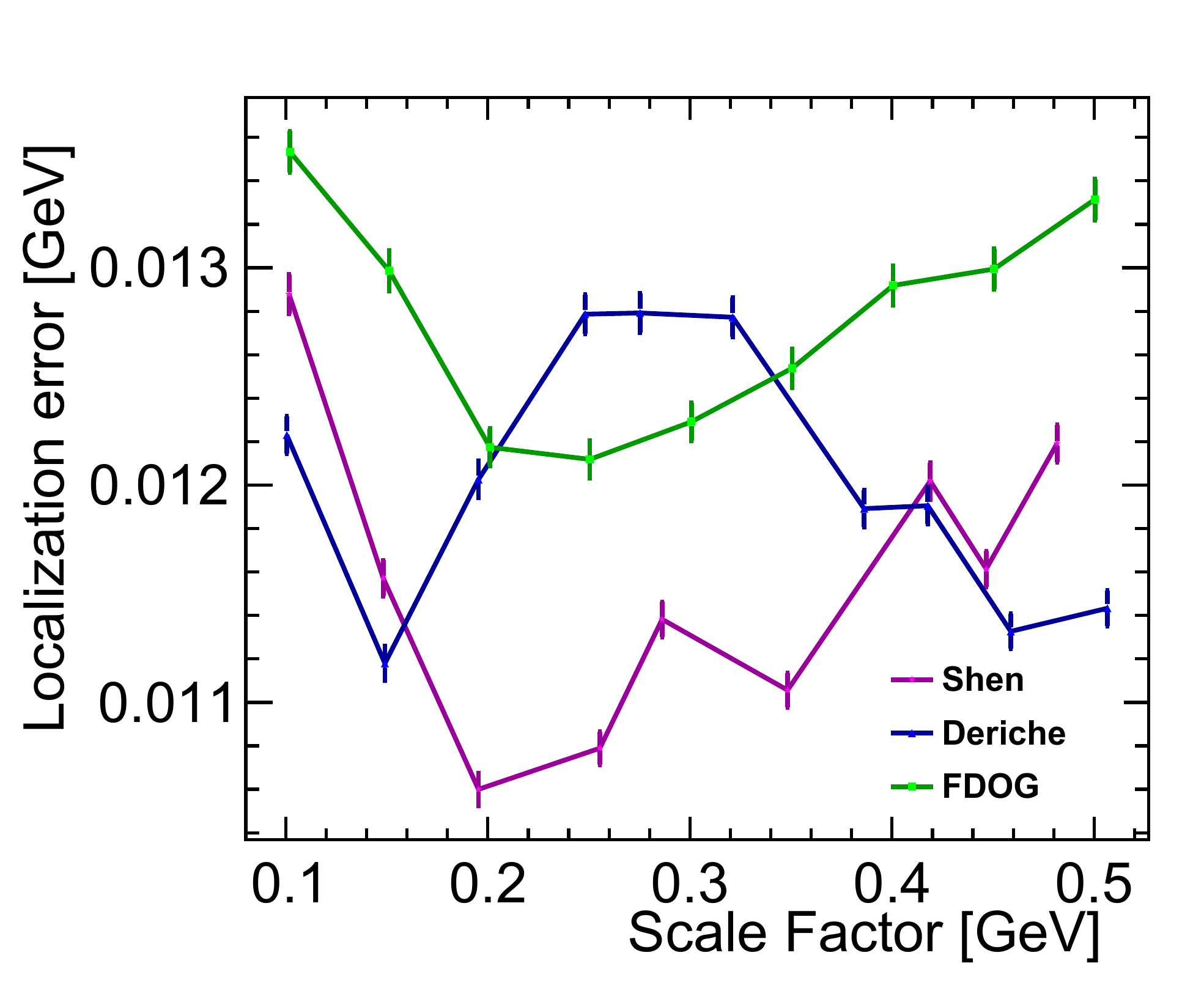}
		\caption[]{Low-momentum edge optimization}
		\label{fig:PMin_LocOptimization}
	\end{subfigure}
	\caption[Edge finder optimization]{The plots show the filter optimization graphs for the FDOG, Deriche and Shen filters, applied to the histograms shown in fig.~\ref{fig:PMax_FinalSpectrum} and fig.~\ref{fig:PMin_LocOptimization}, on the left and on the right respectively.}
	\label{fig:FinalLocOptimization}
\end{figure}

To optimize the parameters of the edge detection algorithm we performed the procedure described in Sec.~\ref{sec:edge-optimization}. The  distributions shown in Figs.~\ref{fig:PMax_FinalSpectrum} and~\ref{fig:PMin_FinalSpectrum} are used as a template for a toy Monte Carlo with which we optimized the binning of the distribution and minimized the filter localization error $\bm{\sigma_L}$. As can be seen from Figs.~\ref{fig:PMax_FinalSpectrum} and~\ref{fig:PMin_FinalSpectrum}, bin widths of \unit[50]{MeV} and \unit[250]{MeV} for the high-momemtum and the low-momentum edges, respectively, correspond to an average bin-by-bin difference of 25 entries in the edge region, i.e.\ to SNR = 5. 
The plots in Fig.~\ref{fig:FinalLocOptimization} show the localization error of the algorithm as a function of the filter size for this binning. In case of the high-momentum edge, Fig.~\ref{fig:PMax_LocOptimization}, the FDOG filter performs best, and for the final result we use it with a scale factor of \unit[1.2]{GeV}. Larger scale factors could have improved the statistical uncertainty a little further, but, as we will see in the following, the systematic uncertainty on the localisation bias is of a similar size, and thus we to not push the statistical precision to the absolute minimum. Due to its steepness, the low-momentum edge shows a significantly less stable situation in Fig.~\ref{fig:PMin_LocOptimization}. The best performance is reached with the Shen filter,
which is not surprising since it has been designed specifically for the case of very narrow edges. However, the FDOG
performs worse only by about 10\%, relatively, and shows a more stable behaviour. Therefore and in view of the above mentioned localisation bias uncertainty, we chose to use the same filter for all cases and thus employ the FDOG with a scale factor of \unit[0.25]{GeV} for the low-momentum edge final results.
Using these parameters, we measure the high-momentum edge position to be $E^{max} = \unit[98.748 \pm 0.043]{GeV}$ and the low-momentum edge position $E^{min} = \unit[7.409 \pm 0.012]{GeV}$.

\begin{figure}
	\centering
	\begin{subfigure}[b]{0.49\linewidth}
		\centering
		\includegraphics[width=\textwidth]{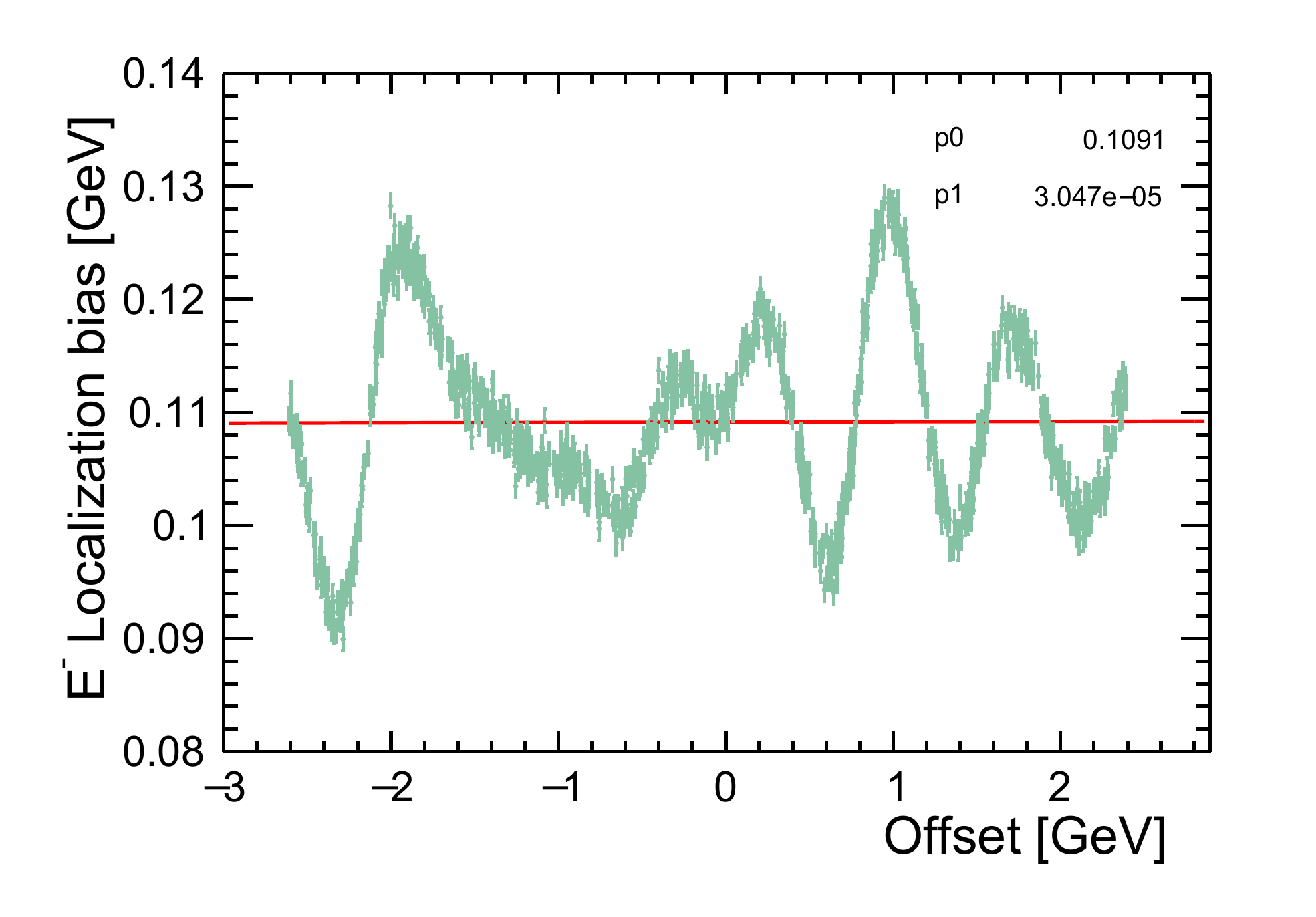}
		\caption[]{Low-momentum edge calibration}
		\label{fig:PMin_BiasCalibration}
	\end{subfigure}
	\hfill
	\begin{subfigure}[b]{0.49\linewidth}
		\centering
		\includegraphics[width=\textwidth]{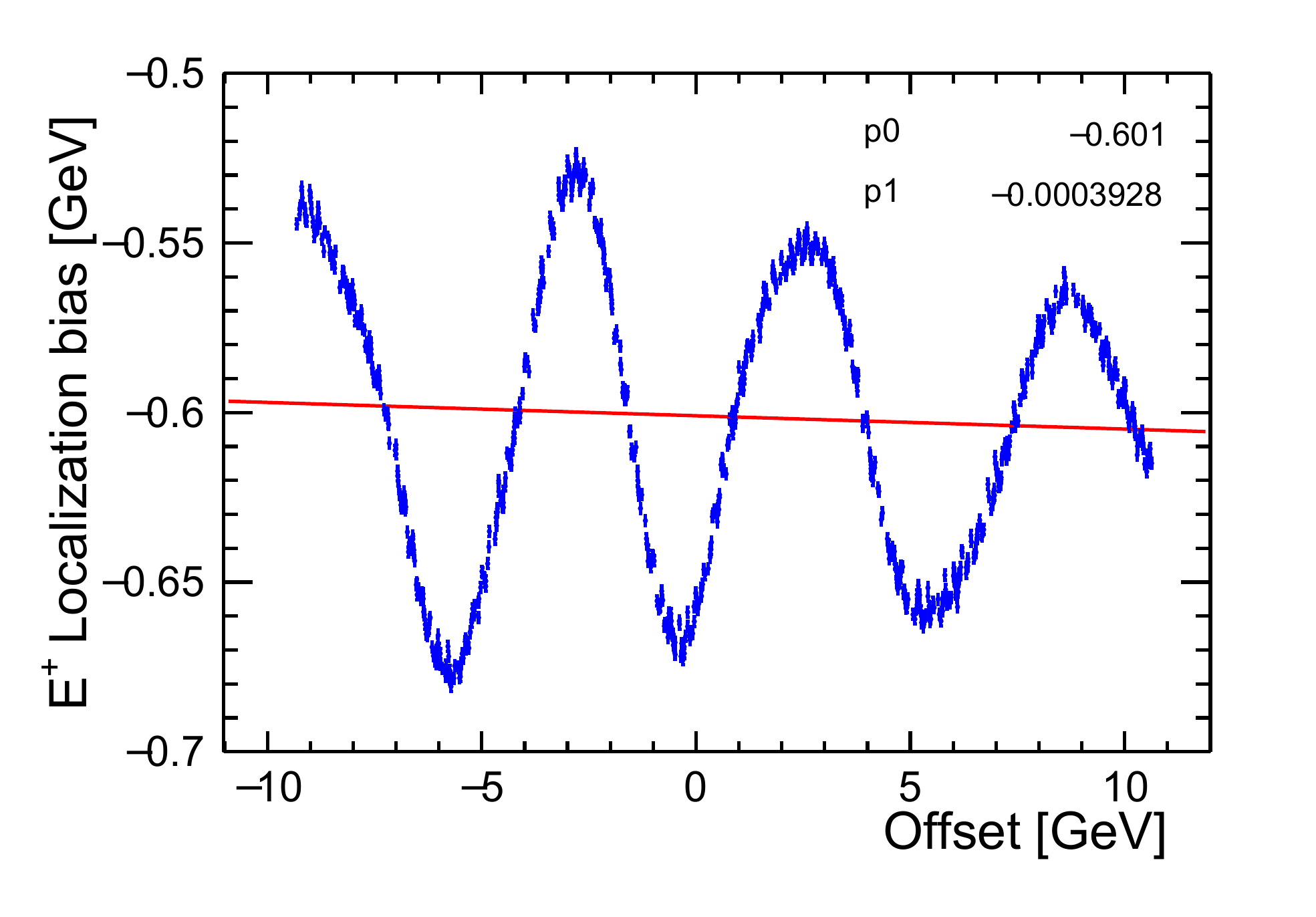}
		\caption[]{High-momentum edge calibration}
		\label{fig:PMax_BiasCalibration}
	\end{subfigure}
	\caption[Calibration graphs of the high- and low-momentum edge measurements]{Calibration graphs of the high-and low-momentum edge measurements. Each data point corresponds to a measurement of the localization parameters on a different template distribution obtained by applying an offset to the signal sample. Superimposed to the measurement set we displayed a first degree polynomial fit to the data showing a long range independence of the bias from the offset.}
	\label{fig:FinalBiasCalibration}
\end{figure}

The second step is the calibration of the edge position measurements, in other words the determination of the localisation bias and its uncertainty. This was done by separating the signal data set from the backgrounds and, while keeping the latter constant, we applied a random offset to the former generating a new template distribution that we used to recalculate the edge position. This procedure is repeated many 
times for different offsets to obtain the distribution of the localization bias as a function of the applied offset. The results of the calibration procedure are shown in Fig.~\ref{fig:FinalBiasCalibration} where the measured bias is displayed as a function of the offset chosen for each measurement. 

A striking feature observed in both plots is a modulation with a constant periodicity. This is caused by the residual effect of the underlying background and is connected with another feature which, albeit not used for the optimization procedure, we defined and characterized in~\cite{Caiazza:2018}. That feature was already described in~\cite{Canny:4767851}, namely the multiple response sensitivity or, otherwise said, the minimum distance of two edges which could be discriminated. In the case of the FDOG filter this distance is roughly equal to 4 times the scale factor of the filter, which is consistent with the periodicity observed in the calibration plots.

Since in an actual measurement one would not know where the true edge position sits within a period, the localisation bias should not be determined directly from the value obtained at an offset of 0, but from the value a linear function fitted to the whole distribution attains at 0 offset. By comparing this value to the ideal edge positions calculated from the generator masses of the supersymmetric particles, the average localization bias is $\bm{\hat{L}}$ \unit[-0.601]{GeV} and \unit[0.109]{GeV} for the two edges respectively. The amplitude of the periodic structure adds a systematic uncertainty to the localisation procedure, which we defined as the standard deviation of all bias values from the calibration procedure, which yields \unit[42]{MeV} for the high-momentum edge and \unit[8]{MeV} for the low-momentum edge. Thus, the systematic uncertainty of the filter method is of the same order or even smaller than the statistical uncertainty on the edge position. Since the periodic structure originates from statistical features in the background, the systematic uncertainty is expected to shrink at a similar rate as the statistical one when considering larger data sets.

\subsection{Edge positions and mass measurements}


Summarizing the results of the edge analysis, we can write our final edge measurements results based on $\unit[500]{fb^{-1}}$ of data with $+80\%$ electron polarization and $-30\%$ positron polarization as:

\begin{align}
\bm{E^{max}} &= \unit[99.349 \pm 0.043 \text{(stat)} \oplus 0.042 \text{(bias)}]{GeV} &\qquad \bm{E^{max}_{\textbf{true}}} &= \unit[99.362]{GeV}\\
\bm{E^{min}} &= \unit[7.300 \pm 0.012 \text{(stat)} \oplus 0.008 \text{(bias)}]{GeV} &\qquad \bm{E^{min}_{\textbf{true}}} &= \unit[7.298]{GeV}
\end{align}

with the \emph{true} positions calculated assuming a nominal centre-of-mass energy of \unit[500]{GeV} as well as 
${M_{\tilde{e}_R} = \unit[126.235]{GeV}}$ and ${M_{\tilde{\chi}^0_1} = \unit[95.586]{GeV}}$ as the \emph{true masses} of the supersymmetric particles.

Finally, we calculate the masses of the selectron and of the neutralino from the measured edge positions, in order to compare them to the results of other studies. As both masses depend on the lower and upper edge, see eqn.~\ref{eq:massesfredesm0}, their calculated values will be correlated. Using a Monte Carlo procedure described in \cite{Caiazza:2018}, we can extract the masses of the $\tilde{e}_R$ and of the $\tilde{\chi}_1^0$ with accuracies of:

\begin{align}
\bm{M_{\tilde{e}_R}} &= 126.20 \pm \unit[0.11]{GeV} &\qquad \bm{M_{\tilde{e}_R}^{\textbf{true}}} &= \unit[126.235]{GeV} \label{eq:SelRFinalMass}\\
\bm{M_{\tilde{\chi}_1^0}} &= 95.56 \pm \unit[0.09]{GeV} &\qquad \bm{M_{\tilde{\chi}_1^0}^{\textbf{true}}} &= \unit[95.586]{GeV} \label{eq:NeuFinalMass}\\
\text{cov}(M_{\tilde{e}_R}, M_{\tilde{\chi}_1^0}) &= \unit[0.0099]{GeV^2} &\qquad \rho &= 0.97
\end{align}

Here, the given uncertainties comprise both the statistical uncertainty on the edge position as well as the uncertainty from the bias correction. They are about a factor of 2 more precise than the best previous results based on the same data set, but obtained using a different method for locating the edge positions as explained in Sec.~\ref{sec:STC4-intro}. The fact that the initial generator values for the two sparticle masses fall within the error range of the measurements is a good indication of the reliability of the measurement algorithm and of the procedure we used to estimate the measurement errors. 

  \section{Edge Measurements with FIR Filters in the Point 5 Study Case}
\label{sec:gauginos}


The second study case, introduced in Sec.~\ref{sec:Point5-intro}, aims to measure the masses of the three lightest gauginos from the kinematic endpoints in $\tilde{\chi}_{1}^{\pm}$ and $\tilde{\chi}_{2}^{0}$ pair production and was carried out with the full ILD simulation \cite{MoradeFreitas:2002kj, AGOSTINELLI2003250} and reconstruction \cite{Marshall:2015rfa, Marshall:2012ry, Thomson:2009rp} chain. In particular, the challenging but frequent hadronic decay modes were considered, which necessitates the full state-of-the-art modelling of the detector as well as full particle flow reconstruction, as argued in~Sec.~\ref{sec:studycases}.   


Considering the two signal reactions presented in equation \ref{eq:point5Reactions}, the event selection searches for topologies characterised by: \textit{(i)} four hadronic jets in the final state, all localised in the main ILD calorimeters, \textit{(ii)} little leptonic activity, and \textit{(iii)} a large amount of missing energy. For both the chargino and the neutralino samples, the selection efficiency is approximately 91\%, however the purity is very low at this stage~\cite{Chera:2018hmr}. This is not only due to the SM background contributions, but also the fact that each signal channel acts as background for the other. To mitigate this, a $\chi^{2}$ test is carried out, comparing the reconstructed di-jet invariant masses to both the \emph{W} and the \emph{Z} boson masses. Based on the obtained $\chi^{2}$ values, $\tilde{\chi}_{1}^{\pm}$ and $\tilde{\chi}_{2}^{0}$ candidates are selected from the events passing the rest of the selection. This enables a final sample purity of 63\% for the charginos and 39\% for the neutralino case at efficiencies of 53\% and 30\%, respectively.

The jet pairing is based on the output of a kinematic fit applied with an equal mass constraint, i.e., the fit does not enforce a specific mass value for the di-jet pairs, but requires that they have equal masses to reflect the expected presence of the gauge boson pair in the final state. For this purpose, the \texttt{MarlinKinFit} \cite{marlinKinfit} package was used. The di-jet configuration providing the largest fit probability was chosen and used further in the analysis. 

\begin{figure}[h]
	\begin{center}
		\begin{subfigure}[h]{0.48\textwidth}
			\begin{centering}
				\includegraphics[width=1.\textwidth]{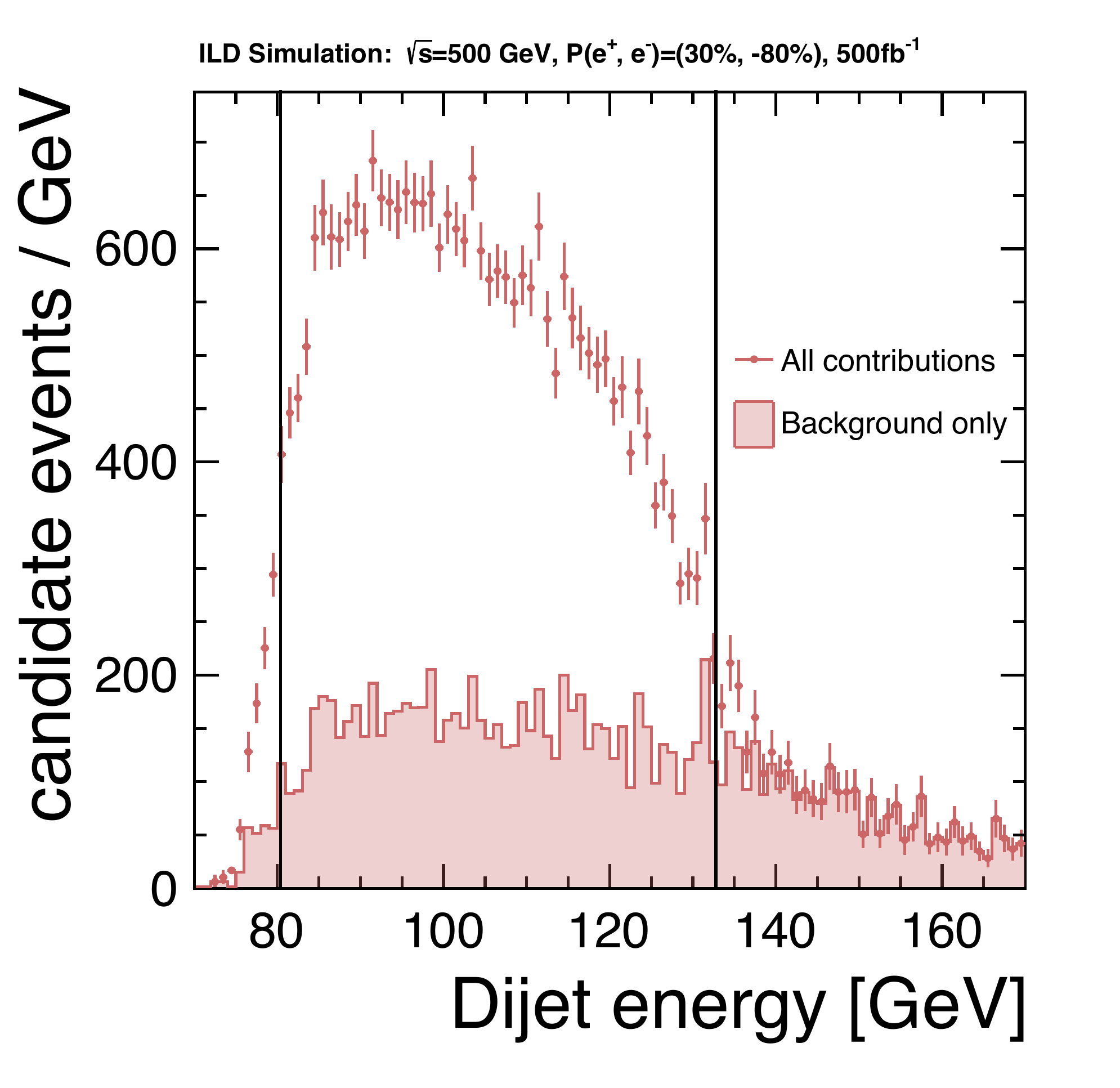}
				\caption{$\tilde{\chi}^{\pm}_{1}$ candidates}
				\label{fig:chi1Observables}
			\end{centering}
		\end{subfigure}\hfill
		\begin{subfigure}[h]{0.48\textwidth}
			\begin{centering}
				\includegraphics[width=1.\textwidth]{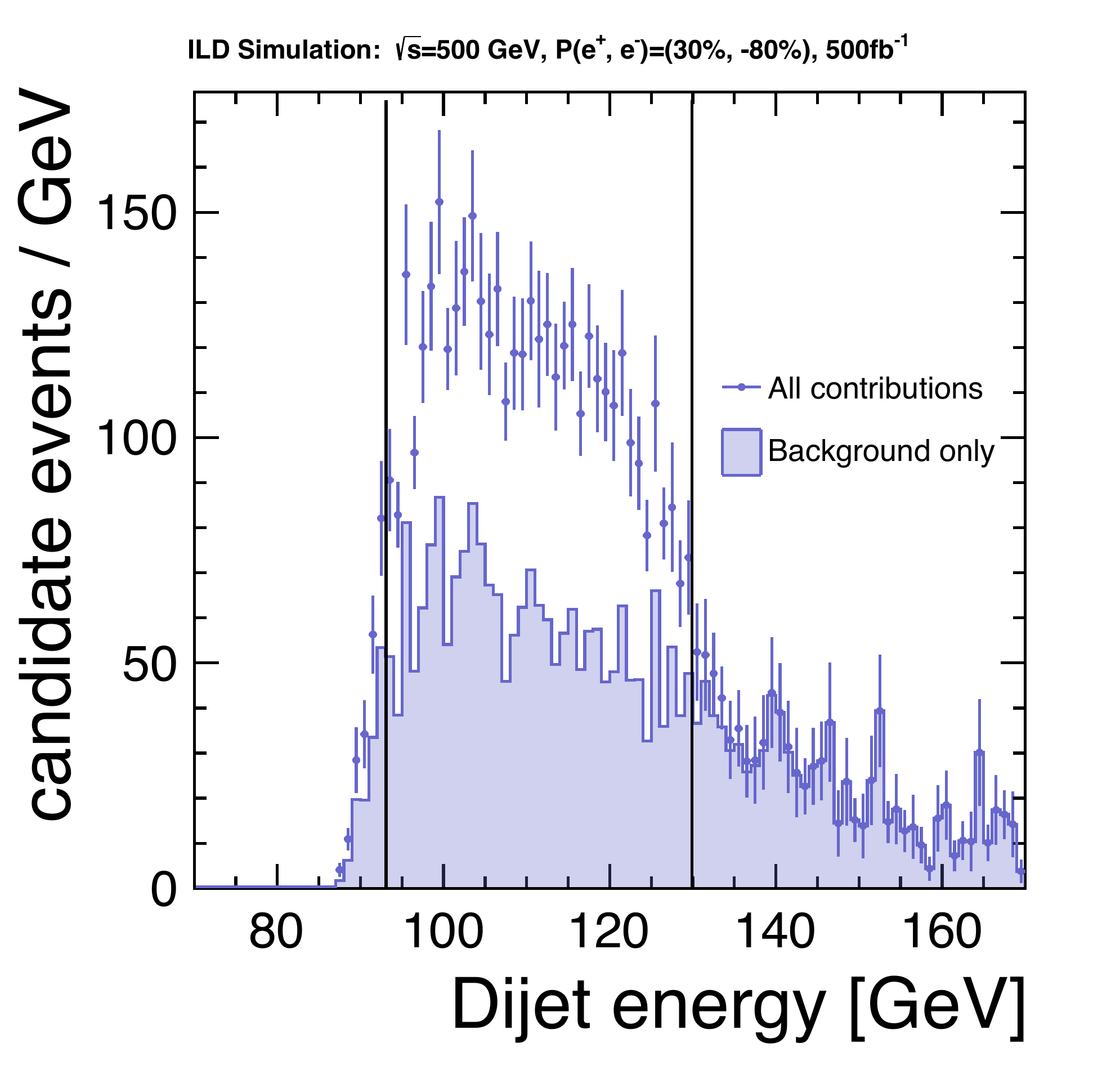}
				\caption{$\tilde{\chi}^{0}_{2}$ candidates}
				\label{fig:chi2Observables}
			\end{centering}
		\end{subfigure}
		\caption{The di-jet energy spectra from (a) the chargino and (b) the neutralino candidate events used in the kinematic edge extraction. Distributions based on events simulated with a detailed description of the ILD detector concept. The black vertical lines indicate the positions of the analytically computed kinematic edges.}
		\label{fig:p5Observables}
	\end{center}
\end{figure}

The relevant observable for determining the gaugino masses is the di-jet energy. The two distributions of this observable, obtained from the sets of $\tilde{\chi}_{1}^{\pm}$ and the $\tilde{\chi}_{2}^{0}$ candidate events, after applying the kinematic fit, are shown in Fig.~\ref{fig:p5Observables}.  
Based on the masses (\emph{M}$_{\tilde{\chi}_{1}^{\pm}}$, \emph{M}$_{\tilde{\chi}_{2}^{0}}$ and \emph{M}$_{\tilde{\chi}_{1}^{0}}$) assumed in the Point 5 model, the ideal positions of the kinematic edges can be calculated analytically. They are indicated by the vertical lines in Fig.~\ref{fig:p5Observables}. This calculation however does not take into account important aspects like e.g.\ the natural width of the decaying gauge bosons, the shape of the beam energy spectrum or the finite resolution of the detector and the reconstruction algorithms, which lead to a much more complicated shape of the realistic spectra, and can even shift the edge positions. All these effects are taken into account in our simulation and are present in the distributions in Fig. \ref{fig:p5Observables}. Therefore, we expect that a calibration procedure will be necessary in order to retrieve the dependency of the edge positions on the particle masses. 

As in the STC4 analysis, an FIR filter with the first derivative of a Gaussian (FDOG) kernel was applied to measure the edge positions in the energy spectra shown in Fig.~\ref{fig:p5Observables}. Before performing the edge measurement, the parameter values of the FIR filter were optimised following the receipe introduced in Sec.~\ref{sec:edge-optimization}.

\subsection{Filter Parameters Optimisation}


For the optimisation, 10$^{4}$ variations of the di-jet energy spectra of the $\tilde{\chi}^{\pm}_{1}$ and the $\tilde{\chi}^{0}_{2}$ candidates were randomly generated from the original ones (Fig.~\ref{fig:p5Observables}) using a toy Monte Carlo approach.

The first step was to choose an appropriate binning of the input data. The di-jet energy spectra upper edges are the noisiest ones not only due to beam energy spectrum effects, but also because the Standard Model background contributions become predominant in that region. Particularly in case of the $\tilde{\chi}^{0}_{2}$ spectrum, shown in Fig.~\ref{fig:chi2Observables}, a SNR > 5 cannot be reached unless the whole upper egde is collapsed into 2-3 bins. With a binning of 1\,GeV per bin, the average count difference between neighbouring bins in the upper edge region is 7.5, corresponding to an SNR $\approx 2.7$, for which only a small penalty in terms of the localization error is expected, c.f.\ Fig.~\ref{fig:FilterSmoothLocvsSNR}. Both parameters, the scale factor of the filter and the binning of the input distribution were varied to verify that the performance expectations from Sec.~\ref{sec:edgedetect} also hold in the realistic physics example.

\begin{figure}
	\begin{center}
		\begin{subfigure}[h]{0.48\textwidth}
	  	   \begin{centering}
		   \includegraphics[width=1.\textwidth]{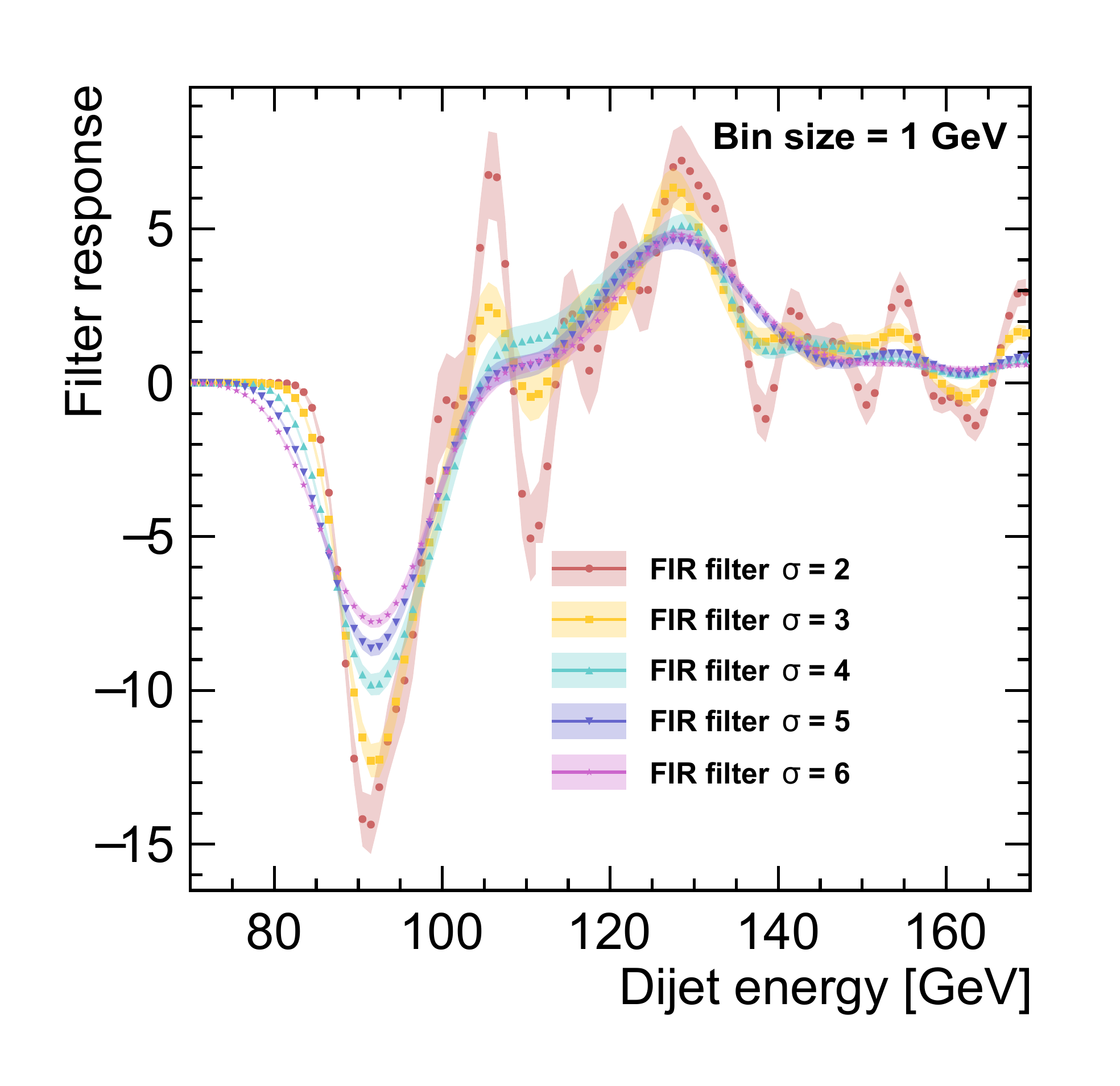}
	       \vspace{-5mm}
	 	   \end{centering}
		\end{subfigure}	
	    ~ 
	   	\begin{subfigure}[h]{0.48\textwidth}
	   	   \begin{centering}
	   	   \includegraphics[width=1.\textwidth]{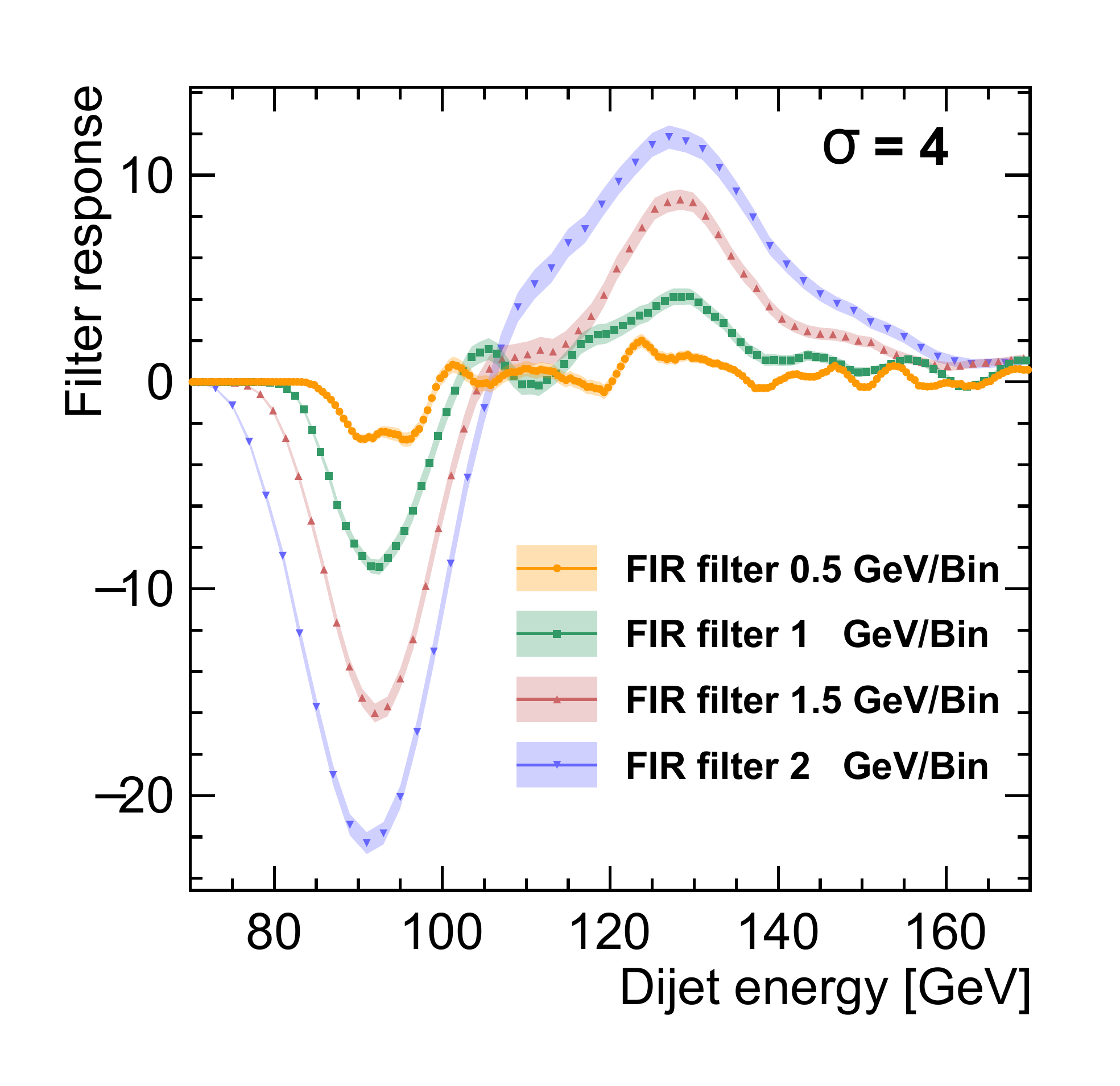}
	   	   \vspace{-5mm}
	   	   \end{centering}
	   	\end{subfigure}   	      
		\caption{Filter responses obtained when varying the Gaussian $\sigma$ parameter (left) and the input data binning (right) in the context of the optimisation procedure. The shown distributions were attained using the simulated $\tilde{\chi}^{0}_{2}$ candidate events. A very similar behaviour was observed in the case of the $\tilde{\chi}^{\pm}_{1}$ candidates.}
		\label{fig:optimisation}
	\end{center}
\end{figure}

\textbf{\textsf{Optimisation of the filter scale factor (Gaussian $\sigma$):}}\\
To determine the optimal value of $\sigma$, the binning of the input distributions were made with 100 bins of width \unit[1]{GeV}. The $\sigma$ value was varied within a range of $\sigma \in \{1 \dots 10\}$, in steps of one. The FDOG filter was then applied with each $\sigma$ value on the toy Monte Carlo produced $\tilde{\chi}^{\pm}_{1}$ and $\tilde{\chi}^{0}_{2}$ di-jet energy spectra. The obtained filter responses are shown in Fig.~\ref{fig:optimisation}.

The very low $\sigma$ values could be immediately excluded since the corresponding FIR filter response was too noisy for a meaningful edge extraction. In contrast, all values above 4 provide almost identical filter responses. Consequently, a value of $\sigma$ = \unit[4]{bins} (shown in green) was chosen as most appropriate for the analysis.




\textbf{\textsf{Optimisation of the bin size:}}\\
To find the optimal bin size for the input di-jet energy spectra, the $\sigma$ parameter was fixed to its previously determined optimal value: $\sigma$ = \unit[4]. A total of four different bin sizes were considered and the filter responses obtained for each of them can be seen in Fig.~\ref{fig:optimisation}.
	
It was determined that, with a binning of \unit[0.5]{GeV/bin}, the filter response could not provide clearly distinguishable and relevant maxima due to too much noise. For bin sizes larger than \unit[1]{GeV/bin}, corresponding to smaller step widths, the peaks marking the edge positions became wider, corresponding to larger localization errors, in accordance with the expectation from Fig.~\ref{fig:FilterLocvsWidth}. Thus, a bin size of \unit[1]{GeV/bin} was chosen for the final analysis.

\subsection{Measured Edge Values}
\label{p5:measurededges}	

\begin{figure}[p!]
\centering
\begin{subfigure}{.68\textwidth}
  \centering
  \vspace{-2mm}
  \includegraphics[width=1.\linewidth]{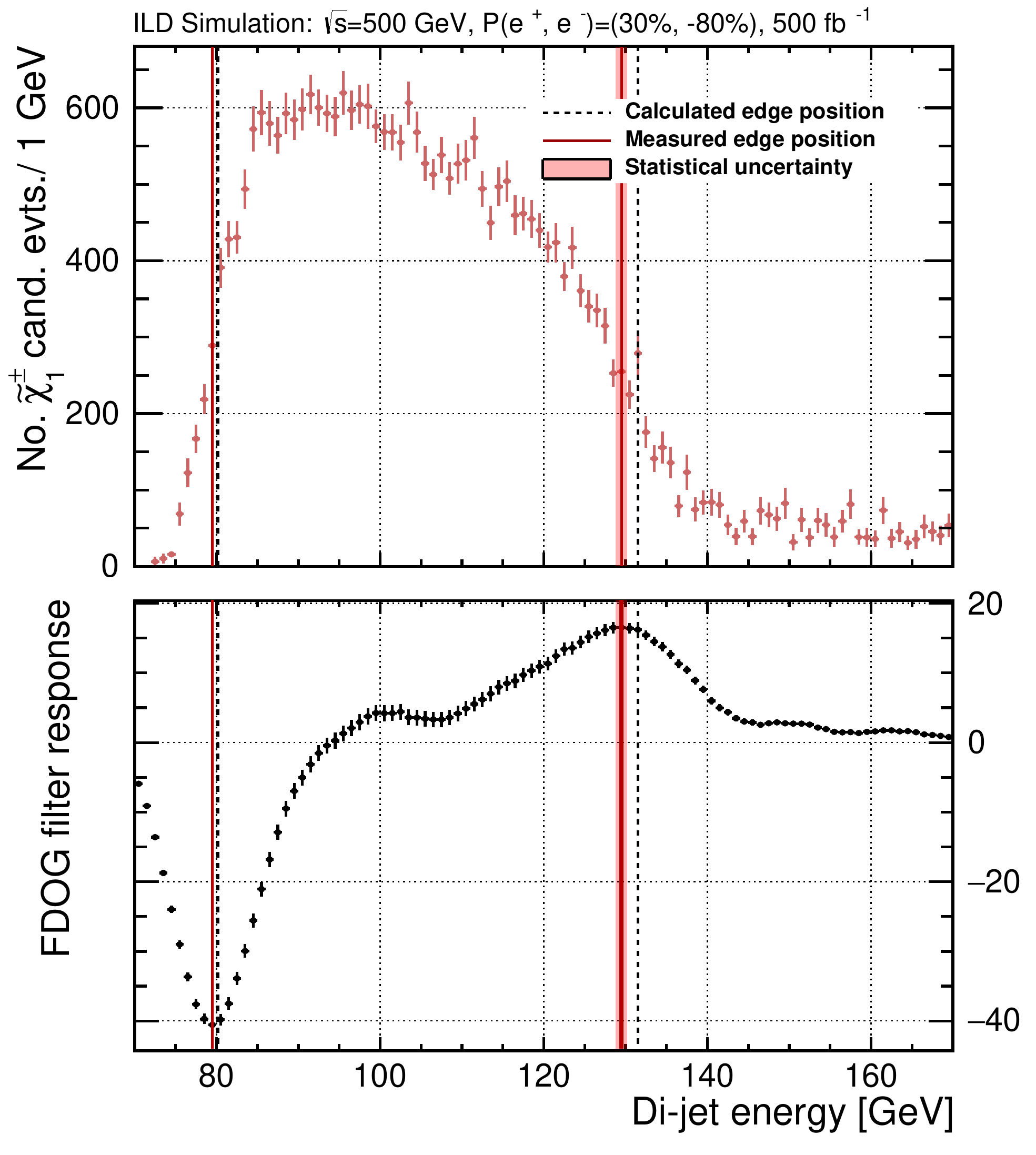}
  \label{fig:sub1}
\end{subfigure}%

\begin{subfigure}{.68\textwidth}
  \centering
  \vspace{-7mm}
  \includegraphics[width=1.\linewidth]{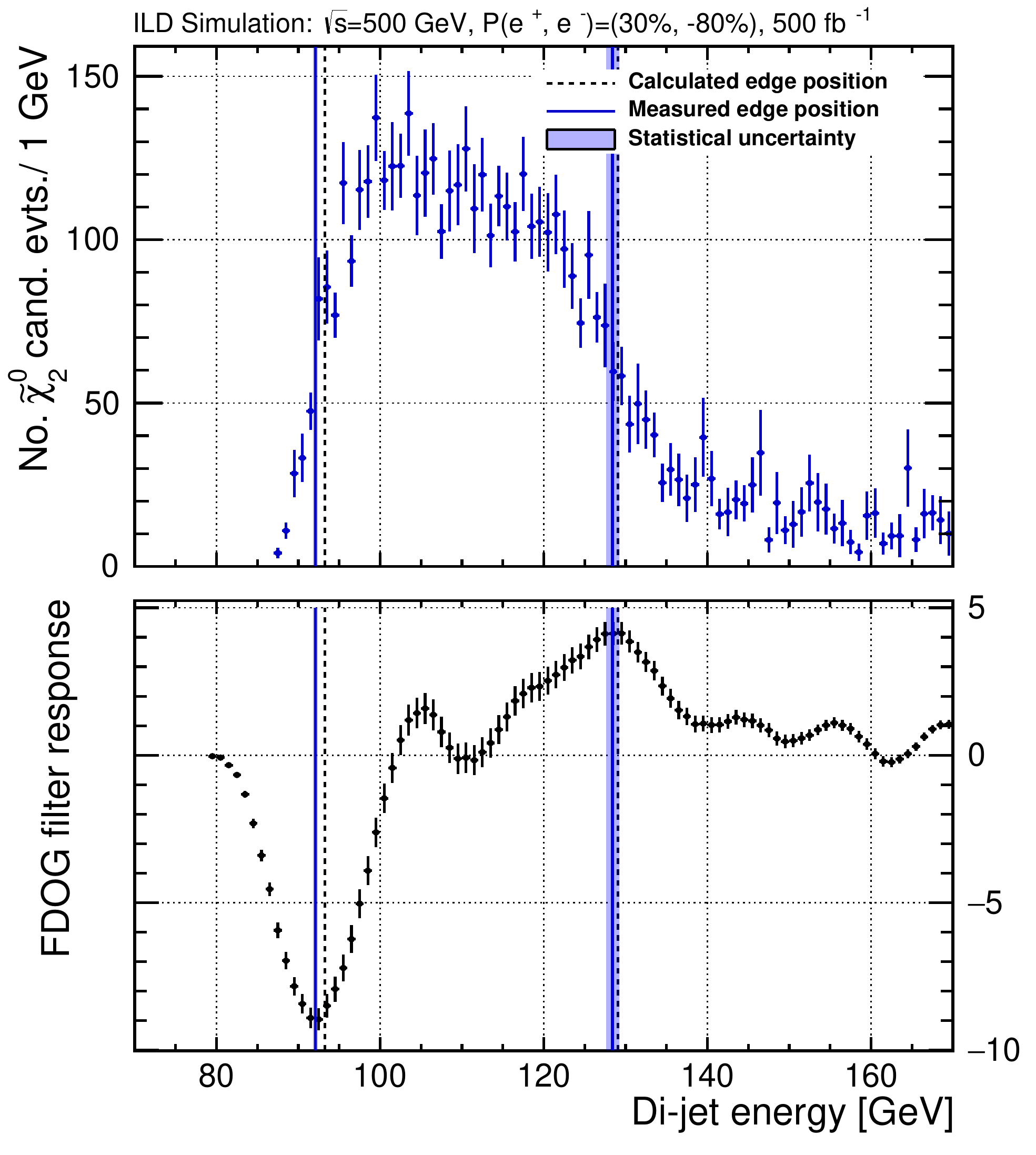}
  \label{fig:sub2}
\end{subfigure}
\vspace{-9mm}
\caption{Results of applying the FIR filter method for extracting the kinematic edge position in the "Point 5" scenario. The model calculated values are also shown for comparison.}
\label{fig:p5Results}
\end{figure}

Following the optimisation process, the filter $\sigma$ parameter was fixed to its determined optimal value, $\sigma$ = \unit[4], and the input data was set to a bin size of \unit[1]{GeV/bin}. The FDOG filter was then applied on the di-jet energy spectra obtained from the $\tilde{\chi}^{\pm}_{1}$ and $\tilde{\chi}^{0}_{2}$ candidate events (Fig.~\ref{fig:p5Observables}) and the filter responses are presented in Fig.~\ref{fig:p5Results}. The upper panels of the figure show the original di-jet energy spectra, i.e. the same distributions from Fig.~\ref{fig:p5Observables}; the lower panels present the computed filter response, i.e. the result of applying the filter on the di-jet energy distributions shown above. The measured edge positions and the corresponding analytically calculated values are indicated by the vertical lines.

The statistical uncertainty was evaluated from 10$^{4}$ toy Monte Carlo experiments based on the di-jet energy spectra for each of the $\tilde{\chi}^{\pm}_{1}$ and the $\tilde{\chi}^{0}_{2}$ candidate samples. The FDOG filter was applied on these spectra and both the low and the high edge positions were determined by performing a Gaussian fit to the maxima present in the filter response. Consequently, four distributions of measured edge values were obtained, one for each of the four edges. The widths (RMS) of these distributions represent the statistical uncertainties.
The central values and uncertainties obtained for $\unit[500]{fb^{-1}}$ at $\sqrt{s}$=\unit[500]{GeV} with $\mathcal{P}$($e^{-}$, $e^{+}$) = ($+80\%$, $-30\%$) are shown in Table~\ref{table:FIRtoyMC}, along with the calculated edge positions.
There is a bias of up to \unit[2]{GeV} in the determination of edge positions, due to the inevitable effects from beam energy spectrum and gauge boson width introduced above, which demonstrates the need for a calibration procedure.

\begin{table}[h!]
	\centering
	\begin{tabular}{|c|c|c|c|c|}
		\hline 
		Edge & $\tilde{\chi}^{\pm}_{1}$ low & $\tilde{\chi}^{\pm}_{1}$ high & $\tilde{\chi}^{0}_{2}$ low &  $\tilde{\chi}^{0}_{2}$ high \\ 
		\hline 
		\hline
		Model calculated & 80.17 & 131.53 & 93.24 & 129.06 \\
	    FDOG filter & 79.49 $\pm$ 0.15 & 129.53 $\pm$ 0.74 & 92.11 $\pm$ 0.31 & 128.38 $\pm$ 0.75\\ 	    
		\hline
	\end{tabular}
	\caption{Kinematic edge positions measured with the FIR filter method, using the first derivative of a Gaussian as the filter kernel. The analytically computed edges are also provided - they represent only an estimation of the real edge positions, as the calculation does not include beam spectrum and gauge boson widths effects.}
	\label{table:FIRtoyMC}
\end{table}

\subsection{Edge Calibration}
The mass calculation presented in Sec. \ref{sec:studycases} entails some approximations. Firstly, the effects of the beam energy spectrum on the edge localisation are not taken into account. In addition, in the Point 5 scenario, the gauge bosons' natural widths also impact the edge positions. This effect is also not considered in the calculation. In order to account for these approximations and also for the reconstruction effects, a mass calibration procedure was performed. However, in this study, the observables determined directly from the data are the edge positions and not the masses, hence, it is the former that must be calibrated. Five new data samples were simulated and reconstructed for this purpose with the ILD full simulation. 

In producing the new data samples, the mass of the LSP, $\tilde{\chi}^{0}_{1}$, was left unchanged and set to the model value of $M_{\tilde{\chi}^{0}_{1}}$ = \unit[115.7]{GeV}. The masses of the two gauginos, $\tilde{\chi}^{\pm}_{1}$ and $\tilde{\chi}^{0}_{2}$, which are almost identical in the Point 5 scenario, were then varied simultaneously from \unit[210]{GeV} to \unit[225]{GeV} in steps of \unit[3]{GeV}. The Standard Model background events produced with the ILD detector full simulation were left unchanged.

Since the LSP mass is not known a priori, a further calibration procedure should involve fixing the two $\tilde{\chi}^{\pm}_{1}$ and $\tilde{\chi}^{0}_{2}$ masses to their Point 5 model values while the $\tilde{\chi}^{0}_{1}$ mass is varied instead. However, this was not performed in the present analysis and remains the subject of future studies.

\begin{figure}[ht!]
	\begin{center}
		\begin{subfigure}[h]{0.45\textwidth}
			\begin{centering}
				\includegraphics[width=1.\textwidth]{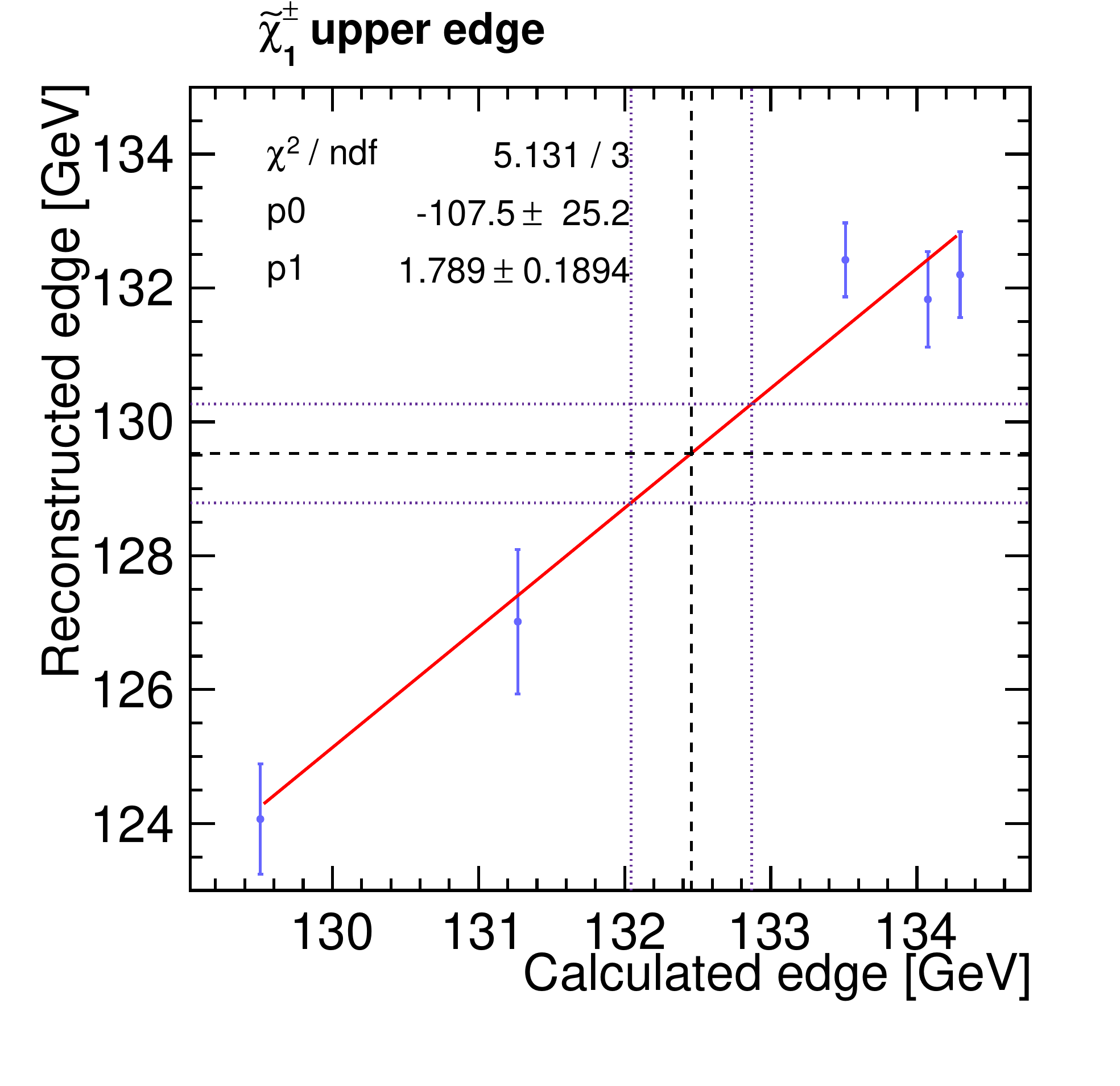}
			\end{centering}
			\vspace{-8mm}
		\end{subfigure}
		\begin{subfigure}[h]{0.45\textwidth}
			\begin{centering}
				\includegraphics[width=1.\textwidth]{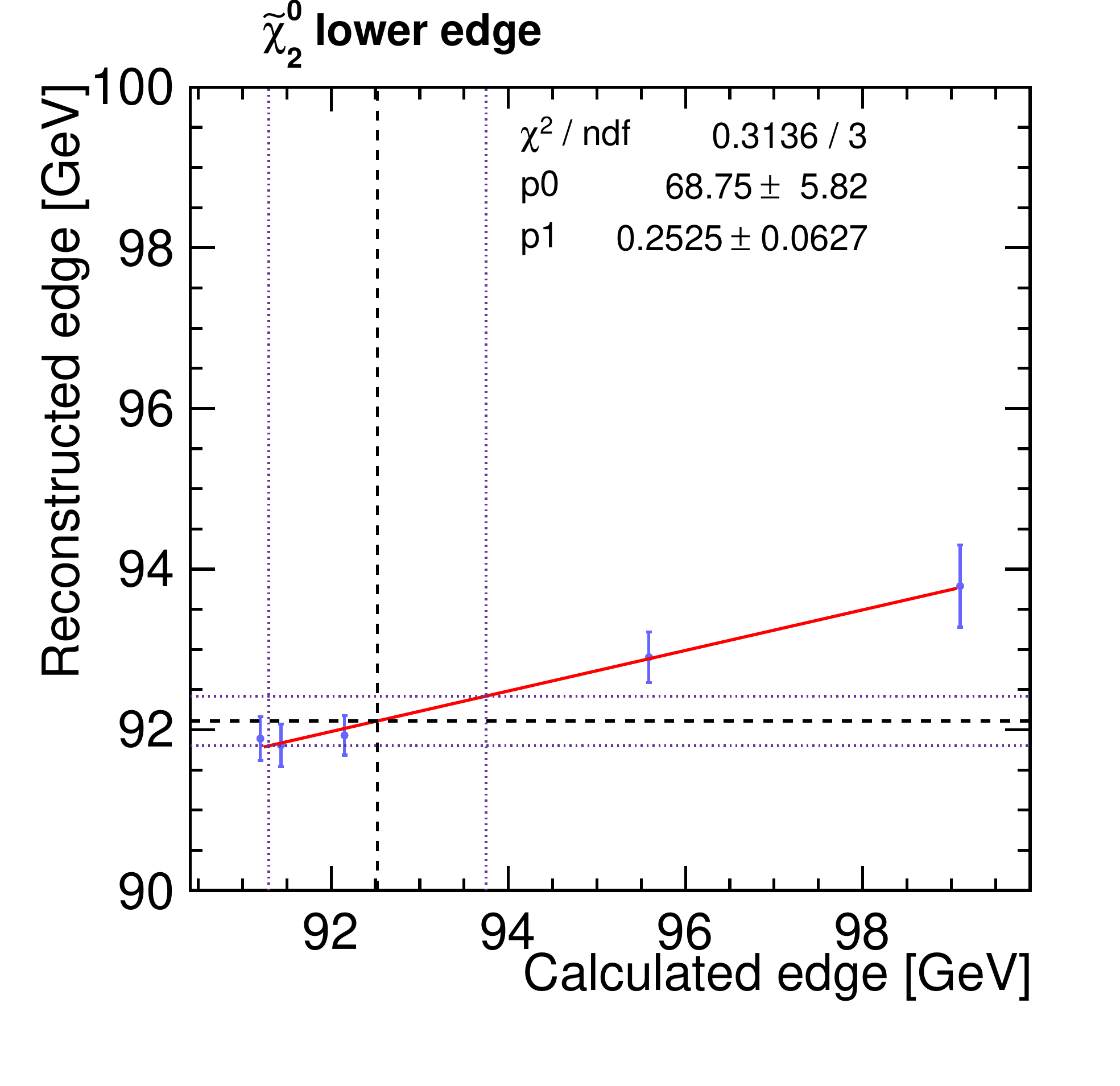}
			\end{centering}
		\end{subfigure}%
		\begin{subfigure}[h]{0.45\textwidth}
			\begin{centering}
				\includegraphics[width=1.\textwidth]{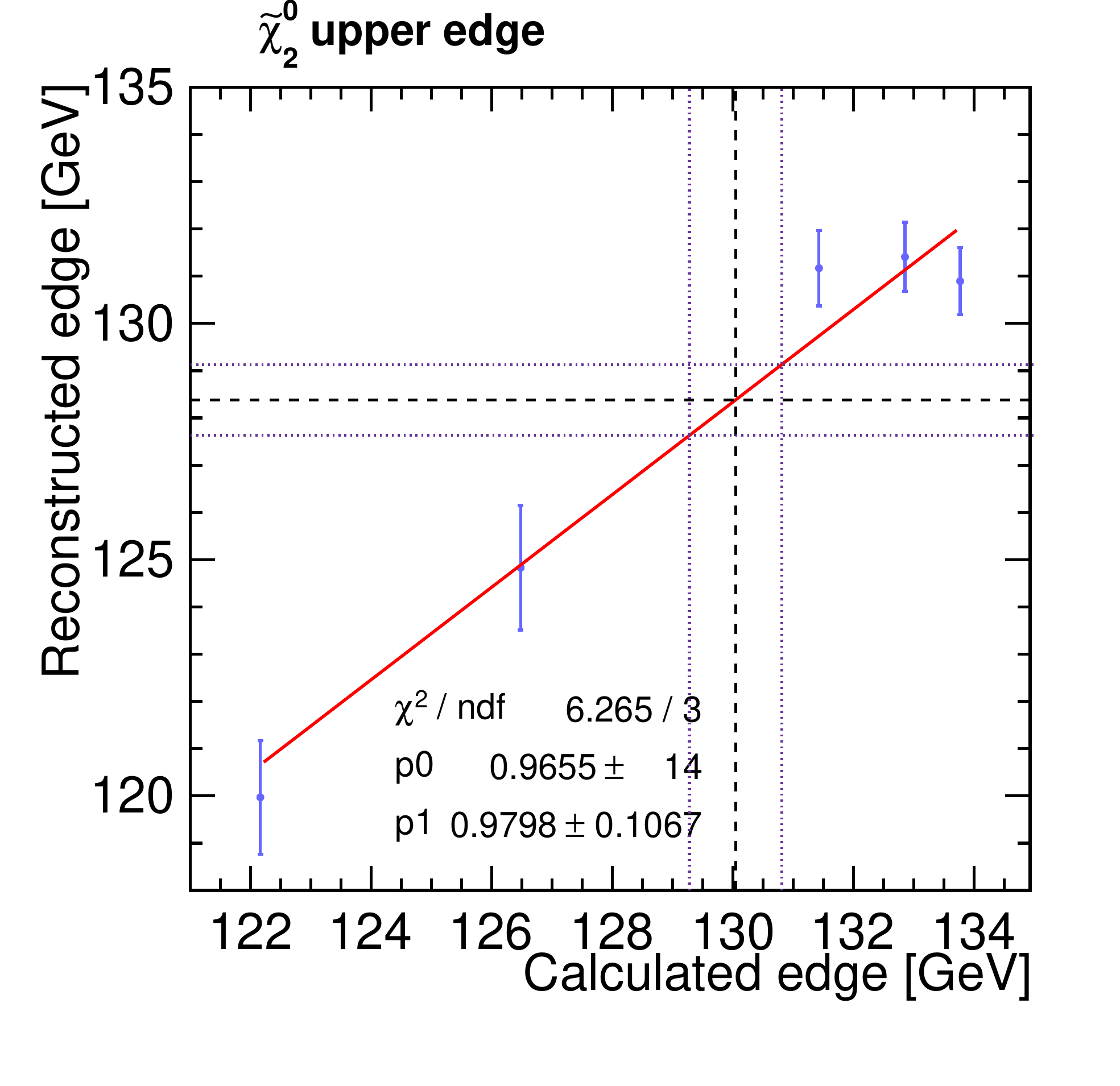}
			\end{centering}
		\end{subfigure}
		\vspace{-8mm}
		\caption{The edge position on reconstruction level versus the model-based calculated values. The red continuous line represents the obtained calibration line. The error projection is also shown (black dashed lines).}
		\label{fig:recVScalc}
	\end{center}
\end{figure}

After producing the five new Monte Carlo data samples, the corresponding \emph{measured} edge positions were obtained by applying the FIR filter method separately. A calibration function was then obtained for each edge, by plotting the corresponding measured values with respect to expected edge values calculated from the input masses. The results are presented in Fig.~\ref{fig:recVScalc}. Since the $\tilde{\chi}^{\pm}_{1}$ low edge overlaps with the mass of the $W$ boson, it is not useful in the mass calculation and was therefore neglected in the calibration procedure.

The ideal slope of the calibration linear fit has a value equal to one. This would indicate that the effects considered in the calibration would have a small impact on the edge positions. While the neutralino upper edge is close to this value, Fig.~\ref{fig:recVScalc} shows a deterioration that can be observed especially in the case of the neutralino lower edge. This is not entirely unexpected since, for the lowest $\tilde{\chi}^{0}_{2}$ mass, $M_{\tilde{\chi}^{0}_{2}}$=\unit[210.7]{GeV}, the phase space volume for the reaction $\tilde{\chi}^{0}_{2} \rightarrow Z \tilde{\chi}^{0}_{1}$ tends to zero ($M_{\tilde{\chi}^{0}_{2}} - M_{\tilde{\chi}^{0}_{1}} \approx$ \unit[95]{GeV}). Furthermore, since the phase space is small, the influence of the beam energy spectrum and of the gauge boson width is much larger. 

To determine each calibrated edge position, the corresponding measured value, presented in section \ref{p5:measurededges}, was projected onto the linear fit (red continuous line in Fig.~\ref{fig:recVScalc}). The intersection point with the calibration line was then further projected onto the $x$-axis. This final projection gives the calibrated value which can be read directly off the horizontal axis. The uncertainty on the measured values was also taken into account in the procedure. 

\begin{table}[ht]
	\centering
	\begin{tabular}{|c|c|c|c|c|c|c|}
		
		\hline
		\hline
		& \makecell{$\tilde{\chi}^{\pm}_{1}$ (high) \\ (\unit[]{GeV})} & \makecell{ $\tilde{\chi}^{0}_{2}$ (low) \\ (\unit[]{GeV})} &  \makecell{$\tilde{\chi}^{0}_{2}$ (high) \\ (\unit[]{GeV})} \\
		\hline
		\hline 
		Reconstructed & 129.53$\pm$0.74 & 92.11$\pm$0.31& 128.38$\pm$0.75 \\
		Calculated    & 132.77          & 93.09 & 129.92 \\
		Calibrated    & 132.46$\pm$0.44 & 92.52$\pm$1.23 & 130.04$\pm$0.77 \\
		\hline     
	\end{tabular}
	\caption{Results of the edge calibration comparing reconstruction level edge values to the calculated values.}
	\label{table:recovscal}
\end{table}  

Table~\ref{table:recovscal} presents the edge calibration results and the comparison between the reconstructed and the calculated edge positions. The cumulative impact of the natural boson width plus the beam and reconstruction effects can be evaluated by comparing the ``Reconstructed'' and the ``Calculated'' columns. It can be seen that, after the calibration, the chargino upper edge decreased by 2\%. In the neutralino case, this effect is of the order of 1\% for the lower edge while the upper edge decreases by 0.7\%. 

Lastly, Table~\ref{tab:massrecvscal} shows the outcome of the calibration procedure when calculating the gaugino masses. While the mean $\tilde{\chi}^{\pm}_{1}$ is shifted lower by 1.2\%, the final measured mass is well compatible with the original model values. The neutralino mass was corrected by 1.6\% while the LSP mass was improved by 2\%. This is, however, achieved at the cost of an  increase of the uncertainty. But most importantly, the main goal of establishing a stable and robust method for the edge localization, which was not available before, has clearly been reached.

\begin{table}[h!]
	\centering
	\begin{tabular}{|c|c|c|c|}
		\hline 
		Gaugino & \makecell{Mass [GeV] \\ (before calib.)} & \makecell{Mass [GeV] \\ (after calib.)}&  Model mass [GeV] \\ 
		\hline 
		\hline
		$\tilde{\chi}^{\pm}_{1}$ & 216.7$\pm$3.1  & 214.1$\pm$4.8 & 216.5 \\ 
		\hline
		$\tilde{\chi}^{0}_{2}$  & 220.4$\pm$1.3 &  216.9$\pm$3.4 & 216.7 \\ 
		\hline
		$\tilde{\chi}^{0}_{1}$  & 118.1$\pm$0.9 &  115.5$\pm$1.8 & 115.7 \\
		\hline
	\end{tabular}
	\caption{Comparison of mass values for the three gaugino masses relevant in the Point 5 scenario before and after the calibration. The model masses are also given for reference.}
	\label{tab:massrecvscal}
\end{table}

  \section{Conclusions} 
\label{sec:conclusions}
In this paper we investigated the application of finite impulse response filters to a typical pattern recognition problem in particle physics, namely the localisation of edges in kinematic distributions. We showed how to choose an appropriate filter kernel and its parameters in presence of statistical fluctuations and irreducible backgrounds in order to optimise the response in terms of efficiency and localisation power. 

Two examples which are highly complementary in signal-to-noise ratio as well as in steepness of the edge have been studied: pair production of selectrons, and the production of charginos and neutralinos, decaying to $W$ and $Z$ bosons, respectively. In case of the selectrons, there is an easy-to-find, very steep edge over a tiny background. The same benchmark has been studied before with various other techniques to extract the selectron and LSP masses.
We showed that with the FIR method, the edge positions can be extracted with a twice better precision than with any other method so far investigated. In case of the charginos and neutralinos, the situation is a priori much more diffcult due to a much worse signal-to-noise ratio and an edge which is smeared out by the detector resolution for hadronic jets -- and by the finite natural width of the intermediate gauge boson. Here, other methods, in particular the fit of an analytic function to the shape of the signal and background parts of the distribution, fail to give reliable results at all. With the FIR filter method, the positions of all four edges could be determined reliably to better than 1\% precision. In both cases, the method was calibrated by varying the input particle masses. After the edge calibration procedure the reconstructed particle masses agree well with their input values. In case of the selectron example, precisions of \unit[110]{MeV} and \unit[90]{MeV} are obtained on the selectron and LSP masses, respectively, assuming a data set of $\unit[500]{fb^{-1}}$ of $e^+e^-$ collisions at a centre-of-mass energy of $\sqrt{s}$=\unit[500]{GeV} and with beam polarisations of $\mathcal{P}$($e^{-}$, $e^{+}$) = (+80\%, -30\%). In case of the gaugino example, the mass of the $\tilde{\chi}^{\pm}_{1}$ was determined with a precision of 2.2\%, while for the two neutralinos precisions of 1.6\% were obtained under the same conditions.
In summary, we showed that the FIR filter technique can be applied very successfully to a variety of typical particle physics distributions. Other physics analysis could benefit from applying this edge detection method.

\section*{Acknowledgements}
This work was supported by the Deutsche Forschungsgemeinschaft (DFG) through the Collaborative Research Centre SFB 676 ``Particles,
Strings and the Early Universe'', project B1.
\printbibliography[title=References]

\end{document}